\documentclass{article}

\usepackage[numbers]{natbib}

\usepackage[preprint]{styles}

\usepackage[utf8]{inputenc} 
\usepackage[T1]{fontenc}    
\usepackage{hyperref}       
\usepackage{url}            
\usepackage{booktabs}       
\usepackage{amsfonts}       
\usepackage{nicefrac}       
\usepackage{amsmath}
\usepackage{microtype}      
\usepackage{xcolor}         
\usepackage{graphicx}       
\usepackage{subcaption}     
\usepackage{caption}        
\usepackage{amssymb}        
\usepackage{wrapfig}        
\usepackage{algorithm, algpseudocode} 

\title{Tree-Conditioned Edit Flows for Ancestral Sequence Reconstruction}

\author{%
  Emil Sharafutdinov \\
  Center for Molecular Protein Science\\
  Lund University \\
  Box 124, 221 00 Lund \\
  \texttt{emil.sharafutdinov@chem.lu.se} \\
  \And
  Ingemar André \\
  Center for Molecular Protein Science \\
  Lund University \\
  Box 124, 221 00 Lund \\
  \texttt{ingemar.andre@chem.lu.se} \\
}

\begin{document}

\maketitle

\begin{abstract}

Ancestral sequence reconstruction (ASR) aims to infer extinct protein sequences at internal nodes of a phylogenetic tree. Classical ASR methods are typically based on continuous-time Markov substitution models, but they treat sites largely independently and handle insertions and deletions only weakly or not at all. We introduce a tree-conditioned edit-flow model for variable-length ASR. Given two descendant sequences and their branch distances to a shared ancestor, the model reconstructs the ancestor through paired bidirectional edit trajectories constrained to agree on a common ancestral state. On a benchmark of experimentally evolved sequences with only context-independent substitutions, the model does not match the accuracy of the best classical method, yet still achieves reasonable performance despite being trained on natural sequences that include insertions, deletions, and substitutions. On a benchmark of natural homologous sequences with abundant insertions and deletions, the model most accurately localizes inferred evolutionary change.

\end{abstract}

\section{Introduction}

Proteins are the result of evolutionary processes that can be traced back to when life first emerged on Earth. The sequences of proteins carry an evolutionary record of this process, but the ancestral states that have given rise to modern proteins are not known. The concept of Ancestral Sequence Reconstruction (ASR), predicting sequences of extinct proteins using statistical models, was introduced in the early 1960s by Zuckerkandl and Pauling \cite{RN5}. ASR has since become a key tool for studying the emergence of protein functions and for protein engineering \cite{RN1, RN4, RN3}.

Classical maximum-likelihood and Bayesian ASR methods \cite{RN7,RN6,huelsenbeck2001empirical} rely on continuous-time Markov substitution models with rate matrices that are fixed or only weakly context-dependent, treating each site as conditionally independent of the surrounding sequence even though real amino acid changes depend on structural, thermodynamic, and functional context \cite{RN9,RN8}. A strength of these methods is that inference is global over the phylogenetic tree: all leaves, including distant outgroups, can inform each internal state through the tree likelihood~\cite{10.1093/sysbio/22.3.240}. However, insertion and deletion operations (indels) are either excluded entirely or ignored in the likelihood calculation, a significant restriction given their functional importance in protein evolution \cite{RN10,RN11,RN13}. Indel-aware models such as ARPIP \cite{10.1093/sysbio/syac050} introduce an explicit insertion--deletion process, but still combine this with site-wise ancestral character reconstruction rather than a sequence-contextual edit model.

More recent neural approaches introduce greater expressivity in modeling of sequence patterns, but do not directly consider phylogenetic trees. AutoregressiveASR \cite{10.1093/molbev/msaf070} processes each branch independently and collapses indels into a 21st token class, conflating indels with substitutions in the modeling. BetaReconstruct \cite{Dotan2026.01.18.700141} treats ASR as a next-token prediction. This approach is alignment-independent and can account for more complex indel evolution, but does not consider tree topology, branch lengths and that ancestors must explain both descendants. With an accurate phylogenetic tree, classical methods substantially outperform BetaReconstruct, suggesting that phylogeny may be highly beneficial for inference of ancestors. 

Here, we introduce \texttt{L\ae{}rad}, a tree-conditioned paired edit-flow model for ancestral sequence reconstruction. L\ae{}rad treats ASR as a branch-conditioned edit process with explicit substitutions, insertions, and deletions; uses paired cross-attention so both descendants inform each child-to-ancestor edit field; and enforces consistency through bidirectional bridge losses and exact-LCA consistency across multiple descendant pairs. At inference time, it decodes candidate ancestors from both descendants under branch-length constraints and reconciles the two directional reconstructions to recover internal nodes recursively across the tree.

\section{Background}

\begin{wrapfigure}{r}{0.50\textwidth}
\centering
\vspace{-2em}
\includegraphics[width=0.40\textwidth]{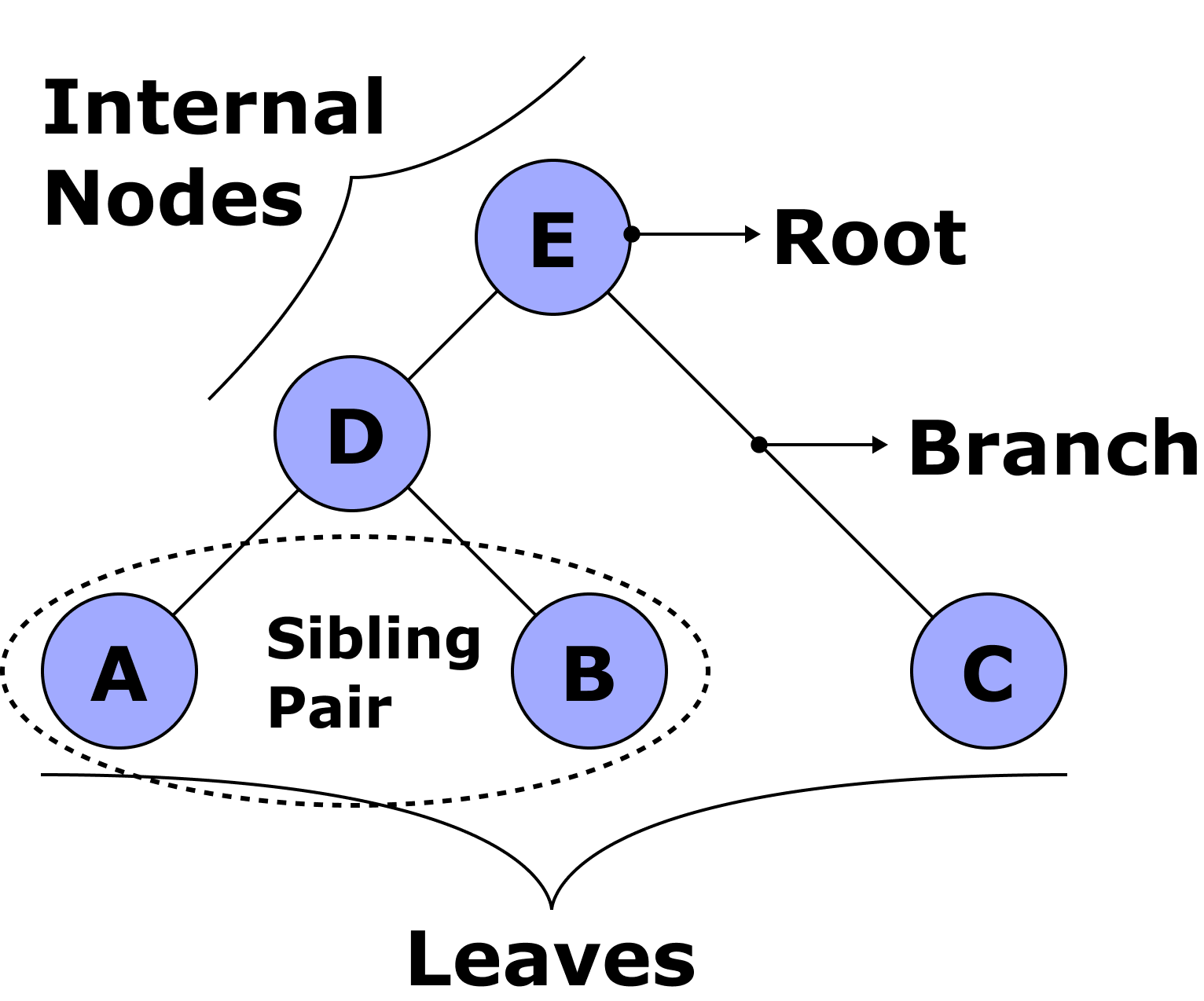}
\caption{A rooted phylogenetic tree with leaves $A$, $B$, $C$ and internal nodes $D$, $E$ (root). Node $D = \operatorname{LCA}(A,B)$ is the deepest shared ancestor of the sibling pair $(A,B)$; $E = \operatorname{LCA}(A,B,C)$ is the root. The length of branches indicate the evolutionary distance between endpoints.}
\label{fig:phylo}
\vspace{-1em}
\end{wrapfigure}

\paragraph{Phylogenetic trees.} Hypothesized evolutionary relationships among a set of biological sequences are represented on a phylogenetic tree. A rooted tree $T=(V,E)$ partitions nodes into observed leaves $L\subset V$ and unobserved internal nodes $I=V\setminus L$. Nodes on the tree are connected by edges. In our implementation, lengths of tree edges are converted into edit-operation budgets by estimating internal states with Fitch parsimony~\cite{b3b975ec-bc62-371a-9f63-3bb7d20f0800} and counting substitutions, insertions, and deletions along each edge. These budgets constrain how much change each reconstruction is allowed to spend; the Fitch states are used only to assign edge distances, not as supervised ancestral targets. For any two nodes $a$ and $b$, their lowest common ancestor $\operatorname{LCA}(a,b)$ is the deepest (from the root) internal node that has $a$ and $b$ as its descendants. In Figure~\ref{fig:phylo}, $D=\operatorname{LCA}(A,B)$ and $E=\operatorname{LCA}(A,B,C)$. 

\paragraph{Ancestral Sequence Reconstruction.}
Classical protein ASR typically starts from a fixed multiple sequence alignment (MSA) of the observed leaf sequences and a phylogenetic tree. At each aligned site \(j\), a continuous-time Markov substitution model computes the probability of the observed leaf states \(x_{L,j}\) by summing over all possible internal states:
\[
p(x_{L,j}\mid T,Q)
=
\sum_{h_{I,j}}
\pi(h_{r,j})
\prod_{(u,v)\in E}
\left[e^{t_{uv}Q}\right]_{h_{u,j},h_{v,j}} ,
\]
where \(h_{v,j}, h_{u,j}\) are the amino-acid states and are unknown at internal nodes (\(u,v \in V\setminus L\)), \(\pi(h_{r,j})\) is the root-state probability under the equilibrium distribution of \(Q\), and \(\left[e^{t_{uv}Q}\right]_{h_{u,j},h_{v,j}}\) is the probability of changing from amino acid \(h_{u,j}\) to \(h_{v,j}\) along branch \((u,v)\) of length \(t_{uv}\). The rate matrix \(Q\) is usually chosen from empirically estimated protein substitution matrices such as JTT \cite{10.1093/bioinformatics/8.3.275}, WAG \cite{10.1093/oxfordjournals.molbev.a003851}, or LG \cite{10.1093/molbev/msn067}. In practice, these models are commonly extended with among-site rate variation, usually through a discrete Gamma distribution over site-specific rate modifiers \cite{10.1093/oxfordjournals.molbev.a040082}; we give the full likelihood in Appendix~\ref{app:classical_asr}.

Given the site-wise likelihoods, ancestral states are commonly inferred either jointly or marginally. Joint reconstruction seeks the assignment of all internal states that maximizes the posterior over the whole tree, and can be solved efficiently by dynamic programming \cite{10.1093/oxfordjournals.molbev.a026369}. Marginal reconstruction instead estimates the most probable state at each internal node and site separately, conditioning on the observed leaves \cite{RN6}. These two procedures optimize different objectives and can disagree at individual positions. More broadly, classical ASR remains strongest when evolution is well described as independent substitutions on a fixed alignment; modeling context-dependent insertions, deletions, and sequence-level edit structure requires a different formulation.

\section{Related Work}

\paragraph{EditFlows.}

Building on recent discrete flow matching methods for discrete generative modeling, edit flows lift flow matching from fixed-length token spaces to variable-length sequences by defining transitions through elementary edit operations \cite{campbell2024generativeflowsdiscretestatespaces,havasi2025editflowsflowmatching}. Let $\mathcal{V}$ denote a discrete vocabulary and let $ \mathcal{X}=\bigcup_{n\geq 0}\mathcal{V}^n $ be the space of all finite sequences over that vocabulary. A continuous-time Markov chain on $\mathcal{X}$ is specified by transition rates $ u_t^\theta(y\mid x)$, where $x,y\in\mathcal{X}$ and $u_t^\theta(y\mid x)$ is nonzero only when $y$ can be obtained from $x$ by one elementary edit operation. The three elementary edits are insertion, deletion, and substitution. Insertions increase sequence length by one, deletions decrease it by one, and substitutions preserve length. This makes edit flows a natural model class for variable-length sequence evolution.

A useful feature of edit flows is that they separate \emph{where} an edit happens from \emph{what} symbol it is emitted. For each operation type $\mathrm{op} \in {\mathrm{ins}, \mathrm{sub}, \mathrm{del}}$ and position $i$, the transition rate factorizes as \[ u_t^\theta(\mathrm{op}(x,i,v)\mid x) =
\lambda^{\mathrm{op}}_{\theta,t,i}(x)\cdot q^{\mathrm{op}}_{\theta,t,i}(v\mid x),
\]
where $\lambda^{\mathrm{op}}_{\theta,t,i}(x)$ is a position- and operation-specific rate and $q^{\mathrm{op}}_{\theta,t,i}(v\mid x)$ is the distribution over the emitted token. The three operations differ in what they produce: substitutions replace position $i$ with token $v$ drawn from $q^{\mathrm{sub}}$; insertions place a new token $v$ from $q^{\mathrm{ins}}$ at position $i$; deletions carry no token distribution — $\lambda^{\mathrm{del}}$ alone determines whether the position is removed.

Deutschmann et al.\ introduce \textit{EvoFlows}, which applies edit flows — substitutions, insertions, and deletions — to single-sequence template-based protein optimisation using a pretrained ESM-2 trunk  \citep{esm2}, showing that the framework recovers target edit distributions and generates natural-like variants across homologous families \cite{deutschmann2026evoflowsevolutionaryeditbasedflowmatching}. We adopt the same backbone and edit-flow view of sequence evolution, but reorient the problem around phylogeny: where EvoFlows optimises a single sequence toward a target distribution, Lærad conditions on a pair of descendants and a known tree, and must produce an ancestral sequence compatible with both children simultaneously under branch-specific edit budgets and topology constraints.

\section{Lærad: Branch-Conditioned Paired Edit Flows for ASR}

L\ae{}rad models ancestral sequence reconstruction as a tree-conditioned edit process on variable-length protein sequences. Given two descendant sequences and their branch distances to a shared ancestor, the model predicts time-dependent substitution, insertion, and deletion rates for both child-conditioned routes rather than directly outputting an ancestor sequence. Training encourages these routes to approach a compatible ancestral region, while inference reconstructs the ancestor by decoding candidate states from both children and selecting the one that best satisfies phylogenetic consistency (Figure~\ref{fig:architecture}).

\subsection{Problem formulation}

Let $T=(V,E)$ be a rooted phylogenetic tree with observed leaves $L \subset V$ and internal nodes $I = V \setminus L$. For a pair of descendant leaf nodes $a$ and  $b$ with sequences $x_a$ and $x_b$ with exact pairwise lowest common ancestor $h=\operatorname{LCA}(a,b)$, let \( d_a > 0\) and \( d_b > 0 \) denote the Fitch-derived branch edit budgets from $x_a$ and $x_b$ to $h$. We interpret these distances as edit budgets and define the expected ancestral interpolation point
\[
\tau = \frac{d_a}{d_a + d_b}
\]
Thus, $\tau$ specifies where the shared ancestor should lie along the descendant-to-descendant bridge, measured in normalized branch-progress coordinates.

L\ae{}rad models ancestral reconstruction as a branch-conditioned continuous-time edit process on variable-length sequences. For a sequence $x \in \mathcal{X}$ and time $t \in [0,1]$, the model outputs position-wise rates for substitution, insertion, and deletion,
\(\lambda^{\mathrm{sub}}_{\theta}(x,t,c),
\lambda^{\mathrm{ins}}_{\theta}(x,t,c),
\lambda^{\mathrm{del}}_{\theta}(x,t,c),\)
together with residue distributions \(
q^{\mathrm{sub}}_{\theta}(\cdot \mid x,t,c)\) and 
\(q^{\mathrm{ins}}_{\theta}(\cdot \mid x,t,c)
\)
where $c$ is the ordered branch condition. For the two views of a descendant pair, we use the ordered branch-budget conditions
\begin{equation}
\label{eq:branch_condition}
c_a =
\left(\frac{d_a}{\bar d}, \frac{d_b}{\bar d}\right),
\qquad
c_b =
\left(\frac{d_b}{\bar d}, \frac{d_a}{\bar d}\right),
\end{equation}
where \(\bar d\) is a dataset-level branch-budget scale. The order identifies the active route, while the scale preserves the approximate distance to the ancestor.

Training is performed on stochastic bridge states sampled between the two descendants. Let
\begin{equation}
\label{eq:bridge}
z_{a\to b}(t) \sim p_t(\cdot \mid x_a, x_b),\qquad
z_{b\to a}(1-t) \sim p_{1-t}(\cdot \mid x_b, x_a)
\end{equation}

denote aligned bridge states sampled on the two opposite routes. L\ae{}rad is trained to predict edit-rate fields on both directions and to regularize the two routes so that they imply a compatible latent ancestral state near $t \approx \tau$. The active objective is
\begin{equation}
\label{eq:objective}
\mathcal{L}
=
w_{\mathrm{base}}\mathcal{L}_{\mathrm{Bregman}}
+
w_{\mathrm{ancestor}}\mathcal{L}_{\mathrm{ancestor}}
+
w_{\mathrm{group}}\mathcal{L}_{\mathrm{group}}.
\end{equation}
where the first term trains local edit dynamics, the second aligns the two opposite-route latent states near the expected ancestral point, and the third enforces consistency across records whose leaf pairs resolve to the same pairwise LCA node. The loss construction is described in Section~4.3, with implementation-level formulas in Appendix~\ref{app:training-objective}.

At inference time, the model does not output an ancestor sequence directly. Instead, it decodes candidate ancestors from both children and selects among the two directional candidates and their consensus merge:
\begin{equation}
\label{eq:inference}
\hat{x}_h
=
\arg\min_{s\in\mathcal{C}(x_a,x_b)}
S(s;x_a,x_b,d_a,d_b).
\end{equation}
The selection score balances branch-budget agreement, compatibility between the two decodes, parsimony, and other fixed inference-time terms:
\begin{equation}
\label{eq:inference_score}
S(s)
=
w_B B(s)
+
w_D D(s)
+
w_P P(s)
+
\mathrm{Reg}(s).
\end{equation}
Here \(B(s)=|\delta(s,x_a)-d_a|+|\delta(s,x_b)-d_b|\) is the branch-budget residual, \(D(s)\) is the disagreement between the two child-conditioned decodes, and \(P(s)=\delta(s,x_a)+\delta(s,x_b)\) is the parsimony cost. The remaining terms are defined in Appendix~\ref{app:inference}.

\subsection{Architecture}

\begin{figure}[h]
    \centering
    \includegraphics[width=1\linewidth]{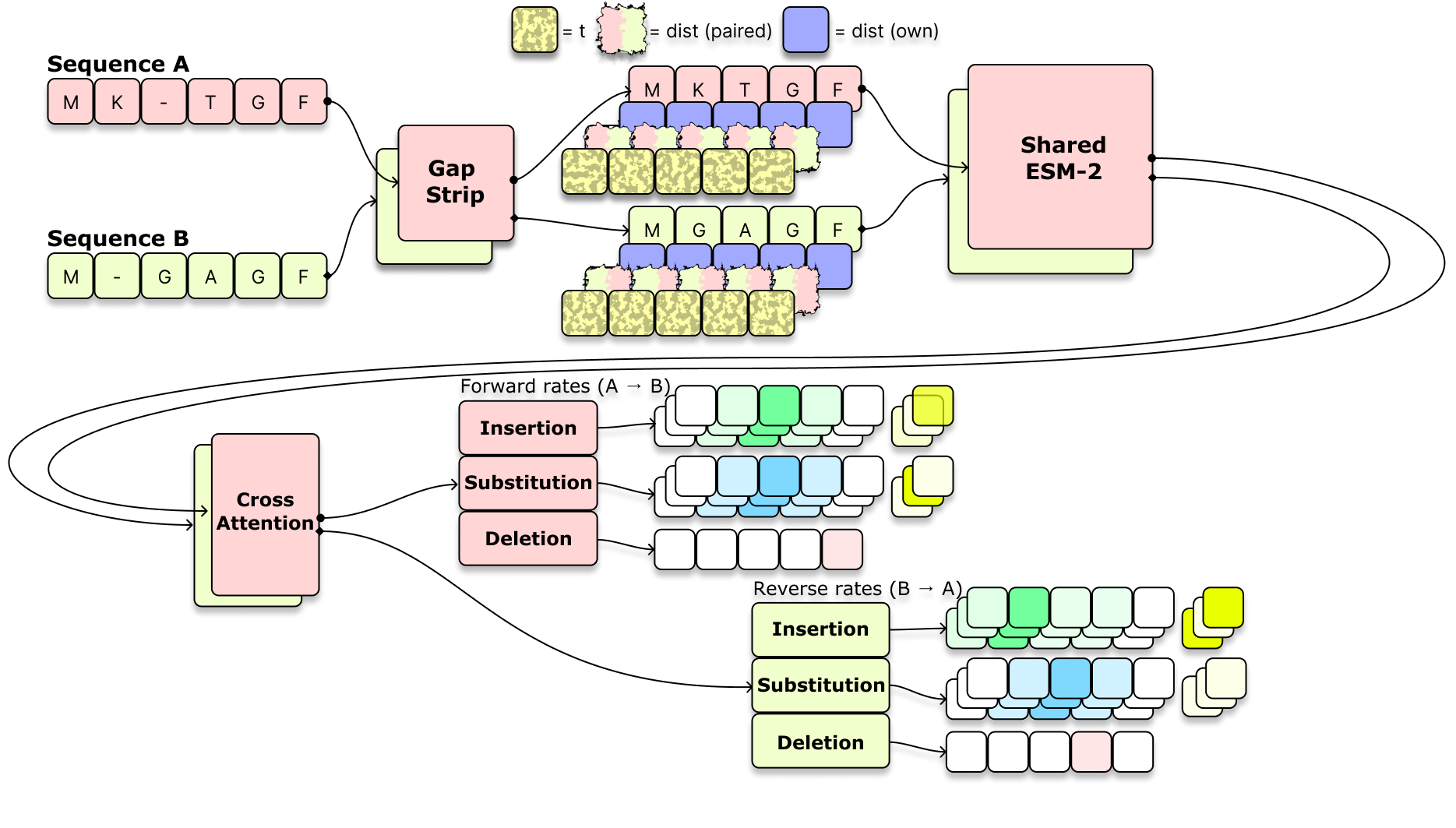}
    \caption{\textbf{Paired Edit-Flow Architecture of L\ae{}rad.}
Aligned descendants \((x_a,x_b)\) are gap-stripped before tokenization, while gap positions are retained for bridge supervision and projection back to alignment coordinates. Each residue receives token, trajectory-time, and ordered branch-budget embeddings. The two budget slots correspond to the active source-to-LCA distance, \(\mathrm{dist(own)}\), and the paired source-to-LCA distance, \(\mathrm{dist(paired)}\); their order is reversed for the two routes. A shared ESM-2 trunk encodes both ungapped sequences, and paired cross-attention fuses descendant context before edit prediction. Operation heads emit positionwise insertion, substitution, and deletion rates; insertion and substitution heads also predict amino-acid distributions, whereas deletion is represented only by rate mass \(\lambda_{\mathrm{del}}\).}
    \label{fig:architecture}
\end{figure}

L\ae{}rad is a paired edit-flow transformer for tree-conditioned ancestral sequence reconstruction. Each training record contains two aligned descendant protein sequences \((x_a,x_b)\), their pairwise LCA identifier, and branch edit distances \((d_a,d_b)\). As shown in Fig.~\ref{fig:architecture}, the aligned descendants are gap-stripped before entering the model, so each decoded route is represented as an ungapped protein-token sequence. At each position, L\ae{}rad adds residue-token, trajectory-time, and ordered branch-condition embeddings. The ordered branch condition identifies the active source sequence and its paired branch budget. The original alignment is retained only for bridge supervision and for projecting predictions back to aligned coordinates during evaluation.

The core sequence encoder is a shared ESM-2 trunk applied to both descendants. After self-attention within each branch, paired fusion layers allow each descendant to attend to the other through cross-attention. This gives the model access to both within-sequence context and paired descendant context before edit rates are predicted.

Output heads then emit positionwise rates for substitution, insertion, and deletion. The substitution and insertion heads additionally predict amino-acid distributions, whereas deletion is represented only by rate mass. The shared trunk provides a common paired protein representation before operation-specific edit fields are predicted.

\subsection{Training}

\paragraph{Batch construction.}
The optimization unit is a descendant-pair ASR record containing two aligned descendants, their pairwise LCA, and branch edit distances \((d_a,d_b)\). During batching, a bridge time \(t\) is sampled and the ordered branch conditions from Eq.~\ref{eq:branch_condition} are constructed for the two bridge orientations. Batches include small groups of pairs whose pairwise LCA is the same exact ancestral node, giving multiple descendant views of that ancestor. This allows different descendant pairs that map to the same exact ancestral node to be constrained toward a shared latent representation, while keeping the optimization unit pairwise. A fixed fraction of each batch is reserved for hard pairs: low-identity pairs, pairs with many edits or indels, long sequences, or pairs from sparsely represented ancestral nodes. This prevents the objective from being dominated by near-identical descendants with nearly unchanged bridge states.

\begin{figure}[h]
    \centering
    \begin{subfigure}{0.49\linewidth}
        \centering
        \includegraphics[width=1.0\linewidth]{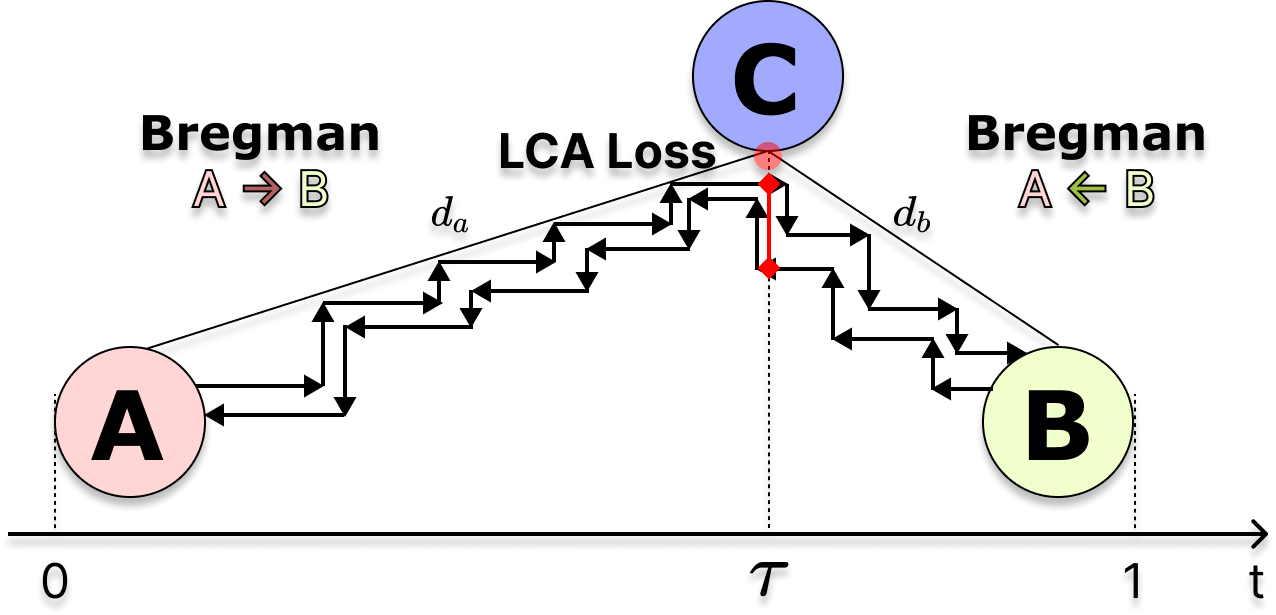}
        \caption{\textbf{Pair reconstruction of parental node}}
        \label{fig:reconstruction_a}
    \end{subfigure}
    \hfill
    \begin{subfigure}{0.49\linewidth}
        \centering
        \includegraphics[width=1.\linewidth]{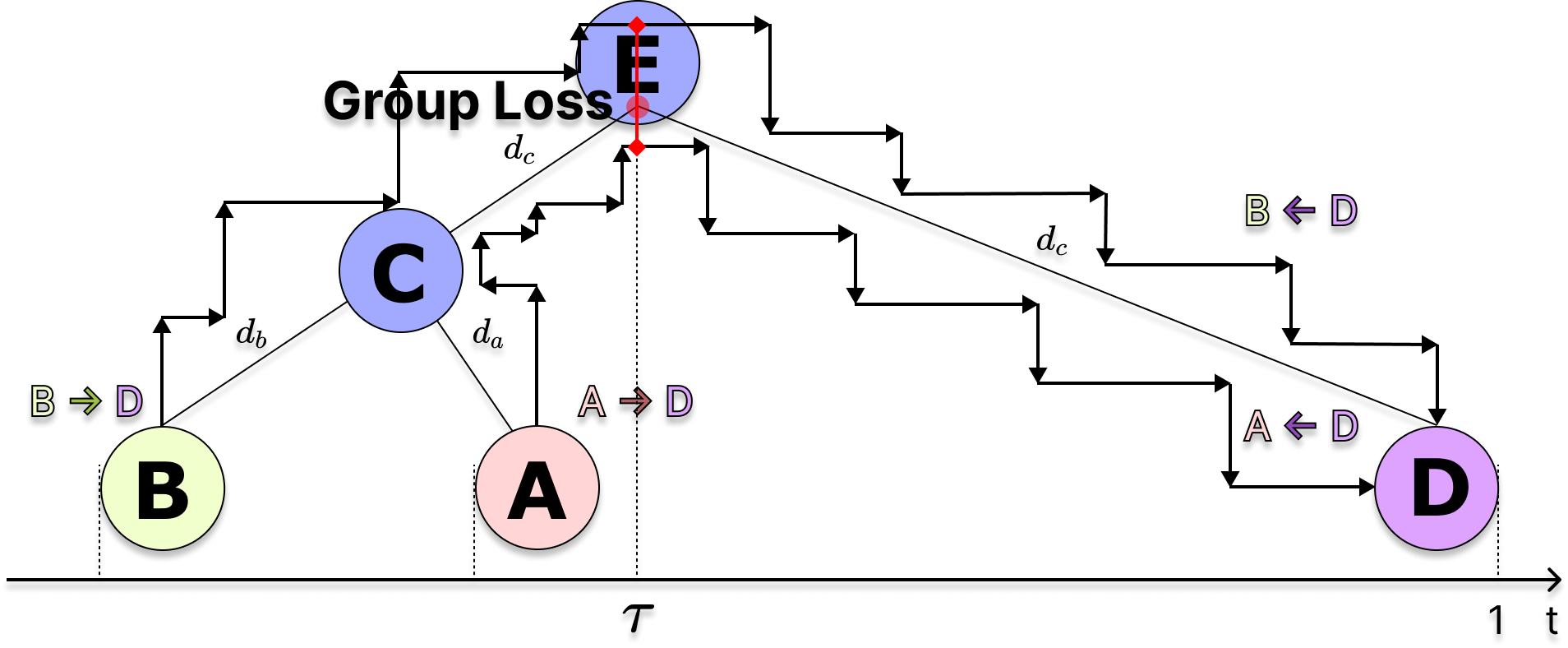}
        \caption{\textbf{Group reconstruction of parental node}}
        \label{fig:reconstruction_b}
    \end{subfigure}
    \caption{
        \textbf{(a) Pair training.} Two descendants (A, B) define a stochastic bridge on $t\in[0,1]$. Each child induces a reverse trajectory toward the other and is supervised by a bidirectional Bregman loss along the full path. The branch-distance ratio $\tau = d_a/(d_a+d_b)$ marks the expected LCA location, where the \emph{LCA loss} becomes critical: it explicitly penalizes disagreement between the two child-conditioned hidden-state trajectories near $t\approx\tau$, forcing them to meet at a common ancestral representation.
        \textbf{(b) Group training.} Two descendant pairs, (A, D) and (B, D), share the same exact LCA node \textbf{E}. Although each pair is trained with its own bidirectional Bregman bridge loss, the \emph{Group loss} compares their mean-pooled latent representations near $t\approx\tau$ and pulls them into agreement. This enforces that different descendant pairs implying the same ancestor converge to a consistent latent state, injecting explicit local tree consistency into training.}
    \label{fig:reconstruction}
\end{figure}

\paragraph{Loss construction.}
The training objective keeps three active signals. Branch lengths are used as conditioning variables, to define the expected LCA location \(\tau=\frac{d_a}{d_a+d_b}\), and later as inference-time selection constraints. For each leaf-pair record, L\ae{}rad samples stochastic bridge states from both descendants and trains the edit-rate field with a bidirectional Bregman loss. This term teaches local edit mechanics: where edits are needed, which operation type is appropriate, and which residue identities are plausible, without using true ancestral sequences.

The second term is an ancestor latent-alignment loss. Near \(t\approx\tau\), valid shared non-gap positions from the two opposite stochastic routes are compared directly in representation space using cosine and RMS/L2 distances. This encourages the two child-conditioned routes to imply a compatible latent ancestral state without forcing token-identical bridge samples.

The third term is group consistency across different descendant-pair views of the same internal ancestral node. Minibatches are sampled so that multiple leaf-pair records may point to the same pairwise LCA, keyed by both family and internal-node identity. Only those records contribute to the group loss; records that merely share a more distant ancestor are not grouped. Near \(\tau\), their mean-pooled latent ancestral representations are pulled together with a combined cosine and RMS/L2 penalty.

Together, these terms instantiate the objective in Eq.~\ref{eq:objective}. The Bregman term trains local edit dynamics, the ancestor alignment term makes the two child-conditioned routes meet in latent space, and the group term makes different descendant pairs that imply the same ancestral node agree on its representation.

\subsection{Inference}

At inference time, \texttt{L\ae rad} reconstructs internal nodes bottom-up on the phylogenetic tree. For a binary internal node with children $(x_a,x_b)$, the branch distances define the expected ancestral bridge location $\tau = d_a/(d_a+d_b)$ and ordered branch conditions $c_a,c_b$ as in Eq.~\ref{eq:branch_condition}. The procedure is summarized in Algorithm~\ref{alg:inference}, where $N$ is the number of sampled trajectories per child.

\begin{algorithm}[h]
\caption{\texttt{Lærad} inference}
\label{alg:inference}
\begin{algorithmic}[1]
\Require Tree $T$, model $\theta$, $N$ samples per child
\Ensure Reconstructed sequence $\hat{x}_v$ for each $v \in I$
\State $\hat{x}_l \leftarrow x_l$ for all $l \in L$ \Comment{initialize leaves}
\For{each $v \in I$ in postorder}
    \State $(\hat{x}_a, \hat{x}_b) \leftarrow \text{children}(v)$ with distances $d_a, d_b$
    \State $\tau \leftarrow d_a/(d_a+d_b)$; form conditions $c_a, c_b$
    \State $\mathcal{A} \leftarrow$ run $N$ flows from $\hat{x}_a$ with $\hat{x}_b$ as context
    \State $\mathcal{B} \leftarrow$ run $N$ flows from $\hat{x}_b$ with $\hat{x}_a$ as context
    \State $(a^*, b^*) \leftarrow \arg\min_{(a,b)\,\in\,\mathcal{A}\times\mathcal{B}}\; S(a,b)$
    \State $m \leftarrow \textsc{Consensus}(a^*, b^*)$ \Comment{position-wise merge}
    \State $\hat{x}_v \leftarrow \arg\min_{s\,\in\,\{a^*,\,b^*,\,m\}}\; S(s)$
\EndFor
\end{algorithmic}
\end{algorithm}

The sets $\mathcal{A}$ and $\mathcal{B}$ contain candidate parent states decoded from $x_a$ and $x_b$ respectively, with the opposite child used as paired context. All pairs $(a,b) \in \mathcal{A}\times\mathcal{B}$ are scored jointly by a weighted sum $S$ combining branch-budget residual, parsimony-style edit cost to both children, pairwise disagreement, indel penalties, and a learned model score. The best-scoring pair $(\hat a, \hat b)$ is then used to construct a merged \texttt{consensus} candidate $m$ by copying matching residues directly and resolving disagreements by local budget compatibility. The final ancestor is selected from $\{ \hat a,\hat b,m \}$ by score $S$, restricted to candidates within budget tolerance if any exist.

\section{Experiments}

We train and evaluate L\ae{}rad using three data sources that play distinct roles: FPbase fluorescent proteins \cite{lambert2016fpbase}, eggNOG orthologous groups \cite{10.1093/nar/gky1085}, and bacteriophage J proteins from the alpha-phage/lambdoid phage dataset \cite{10.1111/evo.13586}.

\subsection{Datasets}

Table~\ref{tab:datasets} summarizes the family-level artifacts used to train and evaluate L\ae{}rad. \texttt{eggNOG} provides the primary large-scale training signal across diverse protein families, branch lengths, and edit patterns. \texttt{FPbase} contributes auxiliary fluorescent-protein pairs of moderate length and supports transfer into fluorescent-protein sequence space. \texttt{Phage J} plays a dual role: it contributes training pairs from the available family artifacts, and it also provides a long, indel-rich evaluation setting through separately benchmarked held-out trees. In addition to the datasets in Table~\ref{tab:datasets}, we evaluate direct ancestral recovery on the experimental fluorescent-protein benchmark of Randall et al.~\cite{randall2016asrbenchmark}, which provides a laboratory phylogeny with known internal ancestral sequences and therefore serves as a real-world ASR benchmark rather than a training source.

\begin{table}[h]
\caption{Training and evaluation datasets. ASR pairs are directed leaf-pair records with an exact LCA and branch edit budgets.}
\centering
\label{tab:datasets}
\begin{tabular}{lrrrl}
\toprule
Dataset & Families & Sequences & ASR pairs & Tree source \\
\midrule
eggNOG  & 5000 & 135{,}356 & 712{,}054 & Precomputed \\
FPbase  & 42   & 955       & 199{,}612 & MAFFT+RAxML \\
Phage J & 2    & 1{,}277   & 104{,}830 & DRYAD FastTree \\
\bottomrule
\end{tabular}
\end{table}

\subsection{Data Preparation}

\paragraph{Sequence filtering and family assignment.}
All datasets are converted into family-level ASR artifacts under a common filtering procedure. A sequence is retained only if it is a valid protein sequence, can be assigned to a family, fits the model context window after alignment, and can be matched to a phylogenetic tree. Protein-level filters remove empty sequences, sequences shorter than 50 or longer than 2000 amino acids, non-standard amino-acid tokens, strong single-residue low complexity, and homopolymer runs longer than 12 residues. For eggNOG, we use families with 2 to 128 members. Aligned records longer than the model context window are discarded. Dataset-specific filtering details are given in Appendix~\ref{app:data_filtering}.

\paragraph{Phylogenetic tree construction and ASR pair generation.}

L\ae{}rad is trained from branch-conditioned relationships between extant sequences rather than from experimentally known ancestral labels. For each retained family, we constructed three necessary components: a phylogenetic tree, a multiple sequence alignment, and branch lengths expressed in edit-operation units. The tree sources used for each dataset are summarized in Table~\ref{tab:datasets}. Dataset-specific details of tree construction, internal-state estimation, and pair generation are deferred to Appendix~\ref{app:tree_pair_generation}.

\subsection{Training Setup}

We evaluate four L\ae{}rad model sizes: Nano (1M parameters), Tiny (3M), Small (9.7M), and Base (37.4M). All variants are trained from scratch on branch-conditioned leaf-pair records with batch size 48 for up to 80{,}000 steps (100{,}000 for Base). The Base model required approximately 5--7 hours on a single NVIDIA A100 GPU; Small required 1--2 hours on a single A100; Tiny and Nano required 1--2 hours on a single NVIDIA RTX 4090. Optimizer and scheduler details are given in Appendix~\ref{app:training_hyperparameters}.

\subsection{Evaluation}

We compare against classical ASR procedures and ASR-capable phylogenetic packages, including Fitch maximum parsimony~\cite{b3b975ec-bc62-371a-9f63-3bb7d20f0800}, PAML/PAML+$\Gamma$~\cite{RN6,10.1093/oxfordjournals.molbev.a040082}, PhyML/PHYLO+$\Gamma$~\cite{guindon2010phyml,oliva2019ambiguity}, ARPIP~\cite{10.1093/sysbio/syac050}, AutoRegressiveASR~\cite{10.1093/molbev/msaf070}, and available IQ-TREE ancestral-state outputs~\cite{nguyen2015iqtree}. For all Lærad variants, reported uncertainty is mean $\pm$ standard deviation over 5 independent runs with different random seeds.

\section{Results}

\paragraph{$\lambda$-phage J-protein ASR with indels.}

The J-protein ID95 benchmark is a long-context ASR setting (70 leaves, 68 internal nodes, 137 scored branches, alignment length \(1263\); ungapped lengths \(1125\)--\(1232\), median \(1137.5\)). We rank methods by observed edit correlation, defined as the Pearson correlation between inferred branch-edit density and empirical leaf-level variation across aligned sites. 

Let \(M^{(m)}_{o,i}\) be the number of branch operations of type \(o\) inferred by method \(m\) at site \(i\), and let \(O_{o,i}\) be the corresponding leaf-vs-reference count, for \(o\in\{\mathrm{sub},\mathrm{ins},\mathrm{del}\}\). Defining \(E^{(m)}_i=\sum_o M^{(m)}_{o,i}\) and \(E^{\mathrm{obs}}_i=\sum_o O_{o,i}\), the primary metric is

\[
\rho_{\mathrm{edit}}^{(m)} = \operatorname{Pearson}\!\left((E^{(m)}_i)_{i=1}^{L},(E^{\mathrm{obs}}_i)_{i=1}^{L}\right).
\]

\textit{Observed indel correlation} uses the same computation but counts only insertions and deletions, so it asks whether indel-rich variation is localized correctly. \textit{Normalized budget error} measures mismatch between inferred edit counts and tree-implied branch budgets, where lower is better. Full metric definitions are given in Appendix~\ref{app:metric_definitions}.

\begin{table}[h]
    \caption{Full-length J-protein ID95 benchmark. Higher is better except for norm. budget error.}
    \vspace{\baselineskip}
    \centering
    \setlength{\tabcolsep}{5pt}
    \begin{tabular}{lrrr}
    \toprule
    Method
    & Obs. edit corr.
    & Obs. indel corr.
    & Normalized budget error \\
    \midrule
    Lærad-Nano
    & 0.773 $\pm$ 0.006
    & 0.684 $\pm$ 0.008
    & 2.092 $\pm$ 0.216 \\
    Lærad-Tiny
    & 0.778 $\pm$ 0.006
    & 0.674 $\pm$ 0.005
    & 2.231 $\pm$ 0.214 \\
    Lærad-Small
    & 0.774 $\pm$ 0.006
    & 0.675 $\pm$ 0.006
    & 1.849 $\pm$ 0.144 \\
    Lærad-Base
    & 0.771 $\pm$ 0.007
    & 0.680 $\pm$ 0.005
    & 1.590 $\pm$ 0.107 \\
    PHYLO-$\Gamma$
    & 0.765
    & --
    & 4.963 \\
    Fitch-MP
    & 0.750
    & --
    & 0.355 \\
    ARPIP
    & 0.740
    & 0.721
    & 1.191 \\
    AutoRegressiveASR
    & 0.739
    & 0.705	
    & 0.234 \\
    IQ-TREE
    & 0.166
    & --
    & 1.932 \\
    \bottomrule
    \end{tabular}
    \label{tab:jprotein-proxy-benchmark}
\end{table}

Table~\ref{tab:jprotein-proxy-benchmark} shows that L\ae rad is strongest on the primary metric: the best observed edit correlation is achieved by L\ae rad-Tiny (\(0.778 \pm 0.006\)), followed closely by L\ae rad-Small (\(0.774 \pm 0.006\)), L\ae rad-Nano (\(0.773 \pm 0.006\)), and L\ae rad-Base (\(0.771 \pm 0.007\)), all above the best non-L\ae rad baseline, PHYLO-\(\Gamma\) (\(0.765\)). This gives L\ae{}rad the highest mean observed edit correlation on this indel-rich long-context benchmark, suggesting stronger localization of inferred evolutionary change across empirically variable sites. IQ-TREE's low observed edit correlation is caused by its gap profile rather than its substitution profile. Its substitution placement is strong, but its aligned-state reconstruction produces a large one-sided deletion signal; since the primary edit metric sums substitutions, deletions, and insertions before correlation, this poorly localized deletion component dominates the aggregate score.

The remaining columns show a narrower picture. Observed indel correlation is strongest for ARPIP (\(0.721\)) and lower for all L\ae rad variants (\(0.674\)--\(0.684\)), indicating that edit-type calibration remains incomplete even when overall edit placement is strong. Budget calibration shows a different pattern: L\ae{}rad-Base has the lowest normalized budget error among L\ae{}rad variants, although several baselines remain substantially better calibrated to branch edit budgets.

Dashes indicate methods for which operation-specific indel correlations are not reported because indels are not explicit model outputs.

\paragraph{Fluorescent-protein phylogeny.}
\begin{table}
  \caption{Fluorescent-protein ASR benchmark. Accuracy is reported over all internal-node sites; lower normalized budget error is better.}
  \vspace{\baselineskip}
  \label{tab:fluorescent-asr}
  \centering
  \begin{tabular}{lrrr}
    \toprule
    Method & Accuracy & Incorrectly inferred sites & Normalized budget error \\
    \midrule
    PHYLO-$\Gamma$ & 97.203 & 107 & 0.374  \\
    ARPIP & 97.098 & 111 & 0.468 \\
    IQ-TREE & 97.020 & 114 & 0.436  \\
    MP & 93.490 & 249 & 0.578 \\
    PAML & 89.987 & 383 & 0.618 \\
    PAML-$\Gamma$ & 89.961 & 384 & 0.459 \\
    AutoRegressiveASR & 87.294 & 486 & 0.631 \\
    Lærad-Nano & 84.397 $\pm$ 0.648 & 596.800 $\pm$ 24.783 & 1.388 $\pm$ 0.084 \\
    Lærad-Tiny & 84.141 $\pm$ 0.762 & 606.600 $\pm$ 29.160 & 1.195 $\pm$ 0.040 \\
    Lærad-Small & 83.216 $\pm$ 0.695 & 642.000 $\pm$ 26.599 & 1.162 $\pm$ 0.081 \\
    Lærad-Base & 82.510 $\pm$ 0.510	 & 669.000 $\pm$ 19.506 & 1.199 $\pm$ 0.137
 \\
    \bottomrule
  \end{tabular}
\end{table}

In addition, we evaluate on the fluorescent-protein benchmark~\cite{randall2016asrbenchmark}, where reference internal-node sequences are available and all methods can be scored by direct sitewise agreement. This benchmark is effectively substitution-only: the reference internal sequences have the same aligned length, indels do not drive the evaluation, and the overall edit burden is low. It is therefore a conservative test for Lærad, because the model is designed for edit operations including insertions and deletions, whereas the strongest baselines are optimized for substitution-based ASR. Results are summarized in Table~\ref{tab:fluorescent-asr}.

As expected, L\ae{}rad trails substitution-specialized methods on this short, low-edit benchmark. Since indels do not drive the task, this is a conservative stress test rather than the setting where edit-flow modeling should be most informative. Among learned methods, AutoRegressiveASR remains stronger on this benchmark (\(87.294\%\) accuracy) than all evaluated L\ae{}rad variants, whose best score is achieved by L\ae{}rad-Nano (\(84.397 \pm 0.648\%\)); both remain well below the strongest classical baselines, led by PHYLO-\(\Gamma\) (\(97.203\%\)) and ARPIP (\(97.098\%\)).

\section{Discussion}

Ancestral sequence reconstruction is fundamentally difficult to evaluate because, in most natural protein families, the true internal ancestral sequences are not observed. Direct ground-truth evaluation is therefore usually impossible, and empirical assessment must rely on the strongest available partial evidence: rare laboratory phylogenies with known ancestors and proxy metrics that test whether inferred mutations are consistent with observable variation in extant sequences. The results in this work should therefore be interpreted as best-supported empirical estimates rather than direct validation against unknown natural ancestors.

On the full-length ID95 J-protein benchmark, L\ae rad achieves the strongest observed edit correlation, suggesting that its inferred ancestral edits are better aligned with empirical site-level variation than those of competing methods. At the same time, operation-specific and budget-aware diagnostics show a narrower picture: observed indel correlation remains below ARPIP, and normalized budget error remains well above the strongest baselines. Within the L\ae rad family, larger models improve branch-budget calibration, although they do not improve the primary ranking metric monotonically.

On the fluorescent-protein benchmark, which is effectively substitution-only, L\ae rad remains below the strongest classical likelihood-based methods and also trails the neural baseline. Taken together, these results suggest that the current strength of L\ae rad lies in edit placement under variable-length, indel-rich reconstruction, while operation-type calibration and branch-budget calibration remain incomplete. 

In exploratory experiments, a larger ESM-2 backbone (650M) reached \(\sim\)89\% accuracy on the fluorescent-protein benchmark despite being undertrained, exceeding the AutoRegressiveASR neural baseline. We do not report it in the main table because it was not trained to the same standard as the reported models, but it suggests that increased scale may improve performance in this substitution-dominated setting.

\section{Conclusion}

We introduced L\ae{}rad, a tree-conditioned paired edit-flow model for ancestral sequence reconstruction that explicitly models substitutions, insertions, and deletions under phylogenetic branch-length constraints. By treating ASR as a paired descendant-to-ancestor edit process rather than as sitewise reconstruction on a fixed alignment, L\ae{}rad extends ancestral inference to variable-length sequence evolution. Empirically, this formulation is most promising in indel-rich settings: on the full-length J-protein benchmark, L\ae{}rad achieves the highest observed edit correlation among evaluated methods, indicating superior localization of inferred evolutionary change. On families where evolution is dominated by substitutions rather than indels, it remains below the strongest classical and neural baselines. Overall, these results support tree-conditioned edit flows as a viable direction for variable-length ASR, while making clear that edit-type and branch-budget calibration remain open problems.

\newpage

{
\small
\bibliographystyle{unsrtnat}
\bibliography{references}

@article{esm2,
  title={Language models of protein sequences at the scale of evolution enable accurate structure prediction},
  author={Lin, Zeming and Akin, Halil and Rao, Roshan and Hie, Brian and Zhu, Zhongkai and others},
  journal={Science},
  volume={379},
  number={6637},
  pages={1123--1130},
  year={2023},
  doi={10.1126/science.ade2574}
}

@misc{havasi2025editflowsflowmatching,
      title={Edit Flows: Flow Matching with Edit Operations}, 
      author={Marton Havasi and Brian Karrer and Itai Gat and Ricky T. Q. Chen},
      year={2025},
      eprint={2506.09018},
      archivePrefix={arXiv},
      primaryClass={cs.LG},
      url={https://arxiv.org/abs/2506.09018}, 
}

@article{randall2016asrbenchmark,
    title={An experimental phylogeny to benchmark ancestral sequence reconstruction},
    author={Ryan N. Randall and Caelan E. Radford and Kelsey A. Roof and Divya K. Natarajan and Eric A. Gaucher},
    year={2016},
    journal = {Nature Communications},
    volume = {7},
    article-number = {12847},
    URL = {https://www.nature.com/articles/ncomms12847},
    DOI = {https://doi.org/10.1038/ncomms12847}
}

@article{lambert2016fpbase,
    title={{FPbase}: a community-editable fluorescent protein database},
    author={Lambert, T.J.},
    year={2019},
    journal = {Nature Methods},
    volume = {16},
    URL = {https://www.nature.com/articles/s41592-019-0352-8},
    DOI = {https://doi.org/10.1038/s41592-019-0352-8}
}

@misc{deutschmann2026evoflowsevolutionaryeditbasedflowmatching,
      title={{EvoFlows}: Evolutionary Edit-Based Flow-Matching for Protein Engineering}, 
      author={Nicolas Deutschmann and Constance Ferragu and Jonathan D. Ziegler and Shayan Aziznejad and Eli Bixby},
      year={2026},
      eprint={2603.11703},
      archivePrefix={arXiv},
      primaryClass={cs.LG},
      url={https://arxiv.org/abs/2603.11703}, 
}

@article{10.1093/sysbio/syac050,
    author = {Jowkar, Gholamhossein and Pe{\v c}erska, Jūlija and Maiolo, Massimo and Gil, Manuel and Anisimova, Maria},
    title = {{ARPIP}: Ancestral Sequence Reconstruction with Insertions and Deletions under the {Poisson} Indel Process},
    journal = {Systematic Biology},
    volume = {72},
    number = {2},
    pages = {307-318},
    year = {2022},
    month = {07},
    issn = {1063-5157},
    doi = {10.1093/sysbio/syac050},
    url = {https://doi.org/10.1093/sysbio/syac050},
    eprint = {https://academic.oup.com/sysbio/article-pdf/72/2/307/50631874/syac050.pdf},
}

@article{10.1093/bioinformatics/btz305,
    author = {Kozlov, Alexey M and Darriba, Diego and Flouri, Tomáš and Morel, Benoit and Stamatakis, Alexandros},
    title = {RAxML-NG: a fast, scalable and user-friendly tool for maximum likelihood phylogenetic inference},
    journal = {Bioinformatics},
    volume = {35},
    number = {21},
    pages = {4453-4455},
    year = {2019},
    month = {11},
    issn = {1367-4803},
    doi = {10.1093/bioinformatics/btz305},
    url = {https://doi.org/10.1093/bioinformatics/btz305},
    eprint = {https://academic.oup.com/bioinformatics/article-pdf/35/21/4453/50721688/bioinformatics_35_21_4453.pdf},
}

@article{b3b975ec-bc62-371a-9f63-3bb7d20f0800,
 ISSN = {00397989},
 URL = {http://www.jstor.org/stable/2412116},
 author = {Walter M. Fitch},
 journal = {Systematic Zoology},
 number = {4},
 pages = {406--416},
 publisher = {[Oxford University Press, Society of Systematic Biologists, Taylor & Francis, Ltd.]},
 title = {Toward Defining the Course of Evolution: Minimum Change for a Specific Tree Topology},
 urldate = {2026-04-15},
 volume = {20},
 year = {1971}
}

@article {Dotan2026.01.18.700141,
	author = {Dotan, Edo and Wygoda, Elya and Schers, Asaf and Lyubman, Iris and Belinkov, Yonatan and Pupko, Tal},
	title = {Ancestral sequence reconstruction using generative models},
	elocation-id = {2026.01.18.700141},
	year = {2026},
	doi = {10.64898/2026.01.18.700141},
	publisher = {Cold Spring Harbor Laboratory},
	URL = {https://www.biorxiv.org/content/early/2026/01/21/2026.01.18.700141},
	eprint = {https://www.biorxiv.org/content/early/2026/01/21/2026.01.18.700141.full.pdf},
	journal = {bioRxiv}
}

@article{10.1093/molbev/msaf070,
    author = {De Leonardis, Matteo and Pagnani, Andrea and Barrat-Charlaix, Pierre},
    title = {Reconstruction of Ancestral Protein Sequences Using Autoregressive Generative Models},
    journal = {Molecular Biology and Evolution},
    volume = {42},
    number = {4},
    pages = {msaf070},
    year = {2025},
    month = {04},
    issn = {1537-1719},
    doi = {10.1093/molbev/msaf070},
    url = {https://doi.org/10.1093/molbev/msaf070},
    eprint = {https://academic.oup.com/mbe/article-pdf/42/4/msaf070/62748976/msaf070.pdf},
}

@article{10.1111/evo.13586,
    author = {Maddamsetti, Rohan and Johnson, Daniel T. and Spielman, Stephanie J. and Petrie, Katherine L. and Marks, Debora S. and Meyer, Justin R.},
    title = {Gain‐of‐function experiments with bacteriophage lambda uncover residues under diversifying selection in nature},
    journal = {Evolution},
    volume = {72},
    number = {10},
    pages = {2234-2243},
    year = {2018},
    month = {10},
    issn = {0014-3820},
    doi = {10.1111/evo.13586},
    url = {https://doi.org/10.1111/evo.13586},
    eprint = {https://academic.oup.com/evolut/article-pdf/72/10/2234/58387831/evolut2234.pdf},
}

@misc{campbell2024generativeflowsdiscretestatespaces,
      title={Generative Flows on Discrete State-Spaces: Enabling Multimodal Flows with Applications to Protein Co-Design}, 
      author={Andrew Campbell and Jason Yim and Regina Barzilay and Tom Rainforth and Tommi Jaakkola},
      year={2024},
      eprint={2402.04997},
      archivePrefix={arXiv},
      primaryClass={stat.ML},
      url={https://arxiv.org/abs/2402.04997}, 
}

@article{10.1093/nar/gky1085,
    author = {Huerta-Cepas, Jaime and Szklarczyk, Damian and Heller, Davide and Hernández-Plaza, Ana and Forslund, Sofia K and Cook, Helen and Mende, Daniel R and Letunic, Ivica and Rattei, Thomas and Jensen, Lars J and von Mering, Christian and Bork, Peer},
    title = {{eggNOG} 5.0: a hierarchical, functionally and phylogenetically annotated orthology resource based on 5090 organisms and 2502 viruses},
    journal = {Nucleic Acids Research},
    volume = {47},
    number = {D1},
    pages = {D309-D314},
    year = {2019},
    month = {01},
    issn = {0305-1048},
    doi = {10.1093/nar/gky1085},
    url = {https://doi.org/10.1093/nar/gky1085},
    eprint = {https://academic.oup.com/nar/article-pdf/47/D1/D309/27437484/gky1085.pdf},
}

@article{RN6,
   author = {Yang, Z. and Kumar, S. and Nei, M.},
   title = {A new method of inference of ancestral nucleotide and amino acid sequences},
   journal = {Genetics},
   volume = {141},
   number = {4},
   pages = {1641-50},
   ISSN = {0016-6731 (Print)
0016-6731 (Linking)},
   DOI = {10.1093/genetics/141.4.1641},
   url = {https://www.ncbi.nlm.nih.gov/pubmed/8601501},
   year = {1995},
   type = {Journal Article}
}

@article{RN11,
   author = {Toth-Petroczy, A. and Tawfik, D. S.},
   title = {Protein insertions and deletions enabled by neutral roaming in sequence space},
   journal = {Mol Biol Evol},
   volume = {30},
   number = {4},
   pages = {761-71},
   ISSN = {1537-1719 (Electronic)
0737-4038 (Linking)},
   DOI = {10.1093/molbev/mst003},
   url = {https://www.ncbi.nlm.nih.gov/pubmed/23315956},
   year = {2013},
   type = {Journal Article}
}

@article{RN3,
   author = {Spence, M. A. and Kaczmarski, J. A. and Saunders, J. W. and Jackson, C. J.},
   title = {Ancestral sequence reconstruction for protein engineers},
   journal = {Curr Opin Struct Biol},
   volume = {69},
   pages = {131-141},
   ISSN = {1879-033X (Electronic)
0959-440X (Linking)},
   DOI = {10.1016/j.sbi.2021.04.001},
   url = {https://www.ncbi.nlm.nih.gov/pubmed/34023793},
   year = {2021},
   type = {Journal Article}
}

@article{RN1,
   author = {Selberg, A. G. A. and Gaucher, E. A. and Liberles, D. A.},
   title = {Ancestral Sequence Reconstruction: From Chemical Paleogenetics to Maximum Likelihood Algorithms and Beyond},
   journal = {J Mol Evol},
   volume = {89},
   number = {3},
   pages = {157-164},
   ISSN = {1432-1432 (Electronic)
0022-2844 (Print)
0022-2844 (Linking)},
   DOI = {10.1007/s00239-021-09993-1},
   url = {https://www.ncbi.nlm.nih.gov/pubmed/33486547},
   year = {2021},
   type = {Journal Article}
}

@article{RN10,
   author = {Savino, S. and Desmet, T. and Franceus, J.},
   title = {Insertions and deletions in protein evolution and engineering},
   journal = {Biotechnol Adv},
   volume = {60},
   pages = {108010},
   ISSN = {1873-1899 (Electronic)
0734-9750 (Linking)},
   DOI = {10.1016/j.biotechadv.2022.108010},
   url = {https://www.ncbi.nlm.nih.gov/pubmed/35738511},
   year = {2022},
   type = {Journal Article}
}

@article{RN4,
   author = {Prakinee, K. and Phaisan, S. and Kongjaroon, S. and Chaiyen, P.},
   title = {Ancestral Sequence Reconstruction for Designing Biocatalysts and Investigating their Functional Mechanisms},
   journal = {JACS Au},
   volume = {4},
   number = {12},
   pages = {4571-4591},
   ISSN = {2691-3704 (Electronic)
2691-3704 (Linking)},
   DOI = {10.1021/jacsau.4c00653},
   url = {https://www.ncbi.nlm.nih.gov/pubmed/39735918},
   year = {2024},
   type = {Journal Article}
}

@article{RN8,
   author = {Pollock, D. D. and Thiltgen, G. and Goldstein, R. A.},
   title = {Amino acid coevolution induces an evolutionary {Stokes} shift},
   journal = {Proc Natl Acad Sci U S A},
   volume = {109},
   number = {21},
   pages = {E1352-9},
   ISSN = {1091-6490 (Electronic)
0027-8424 (Print)
0027-8424 (Linking)},
   DOI = {10.1073/pnas.1120084109},
   url = {https://www.ncbi.nlm.nih.gov/pubmed/22547823},
   year = {2012},
   type = {Journal Article}
}

@article{RN5,
   author = {Pauling, Linus and Zuckerkandl, Emile and Henriksen, Thormod and Lövstad, Rolf},
   title = {Chemical Paleogenetics. Molecular "Restoration Studies" of Extinct Forms of Life},
   journal = {Acta Chemica Scandinavica},
   volume = {17 supl.},
   pages = {9-16},
   ISSN = {0904-213X},
   DOI = {10.3891/acta.chem.scand.17s-0009},
   year = {1963},
   type = {Journal Article}
}

@article{RN9,
   author = {Norn, C. and Andre, I.},
   title = {Atomistic simulation of protein evolution reveals sequence covariation and time-dependent fluctuations of site-specific substitution rates},
   journal = {PLoS Comput Biol},
   volume = {19},
   number = {3},
   pages = {e1010262},
   ISSN = {1553-7358 (Electronic)
1553-734X (Print)
1553-734X (Linking)},
   DOI = {10.1371/journal.pcbi.1010262},
   url = {https://www.ncbi.nlm.nih.gov/pubmed/36961827},
   year = {2023},
   type = {Journal Article}
}

@article{RN7,
   author = {Koshi, J. M. and Goldstein, R. A.},
   title = {Probabilistic reconstruction of ancestral protein sequences},
   journal = {J Mol Evol},
   volume = {42},
   number = {2},
   pages = {313-20},
   ISSN = {0022-2844 (Print)
0022-2844 (Linking)},
   DOI = {10.1007/BF02198858},
   url = {https://www.ncbi.nlm.nih.gov/pubmed/8919883},
   year = {1996},
   type = {Journal Article}
}

@article{RN13,
   author = {Hormozdiari, F. and Salari, R. and Hsing, M. and Schonhuth, A. and Chan, S. K. and Sahinalp, S. C. and Cherkasov, A.},
   title = {The effect of insertions and deletions on wirings in protein-protein interaction networks: a large-scale study},
   journal = {J Comput Biol},
   volume = {16},
   number = {2},
   pages = {159-67},
   ISSN = {1557-8666 (Electronic)
1066-5277 (Linking)},
   DOI = {10.1089/cmb.2008.03TT},
   url = {https://www.ncbi.nlm.nih.gov/pubmed/19193143},
   year = {2009},
   type = {Journal Article}
}

@article{10.1093/molbev/msn067,
    author = {Le, Si Quang and Gascuel, Olivier},
    title = {An Improved General Amino Acid Replacement Matrix},
    journal = {Molecular Biology and Evolution},
    volume = {25},
    number = {7},
    pages = {1307-1320},
    year = {2008},
    month = {07},
    issn = {0737-4038},
    doi = {10.1093/molbev/msn067},
    url = {https://doi.org/10.1093/molbev/msn067},
    eprint = {https://academic.oup.com/mbe/article-pdf/25/7/1307/3520981/msn067.pdf},
}

@article{10.1093/oxfordjournals.molbev.a026369,
    author = {Pupko, Tal and Pe'er, Itsik and Shamir, Ron and Graur, Dan},
    title = {A Fast Algorithm for Joint Reconstruction of Ancestral Amino Acid Sequences},
    journal = {Molecular Biology and Evolution},
    volume = {17},
    number = {6},
    pages = {890-896},
    year = {2000},
    month = {06},
    issn = {0737-4038},
    doi = {10.1093/oxfordjournals.molbev.a026369},
    url = {https://doi.org/10.1093/oxfordjournals.molbev.a026369},
    eprint = {https://academic.oup.com/mbe/article-pdf/17/6/890/65172747/mbev_17_6_0890.pdf},
}

@article{10.1093/oxfordjournals.molbev.a040082,
    author = {Yang, Z},
    title = {Maximum-likelihood estimation of phylogeny from DNA sequences when substitution rates differ over sites.},
    journal = {Molecular Biology and Evolution},
    volume = {10},
    number = {6},
    pages = {1396-1401},
    year = {1993},
    month = {11},
    issn = {0737-4038},
    doi = {10.1093/oxfordjournals.molbev.a040082},
    url = {https://doi.org/10.1093/oxfordjournals.molbev.a040082},
    eprint = {https://academic.oup.com/mbe/article-pdf/10/6/1396/11176354/19YANG.PDF},
}

@article{10.1093/oxfordjournals.molbev.a003851,
    author = {Whelan, Simon and Goldman, Nick},
    title = {A General Empirical Model of Protein Evolution Derived from Multiple Protein Families Using a Maximum-Likelihood Approach},
    journal = {Molecular Biology and Evolution},
    volume = {18},
    number = {5},
    pages = {691-699},
    year = {2001},
    month = {05},
    issn = {0737-4038},
    doi = {10.1093/oxfordjournals.molbev.a003851},
    url = {https://doi.org/10.1093/oxfordjournals.molbev.a003851},
    eprint = {https://academic.oup.com/mbe/article-pdf/18/5/691/23447821/i0737-4038-018-05-0691.pdf},
}

@article{10.1093/bioinformatics/8.3.275,
    author = {Jones, David T. and Taylor, William R. and Thornton, Janet M.},
    title = {The rapid generation of mutation data matrices from protein sequences},
    journal = {Bioinformatics},
    volume = {8},
    number = {3},
    pages = {275-282},
    year = {1992},
    month = {06},
    issn = {1367-4803},
    doi = {10.1093/bioinformatics/8.3.275},
    url = {https://doi.org/10.1093/bioinformatics/8.3.275},
    eprint = {https://academic.oup.com/bioinformatics/article-pdf/8/3/275/479648/8-3-275.pdf},
}

@article{huelsenbeck2001empirical,
  title={Empirical and Hierarchical Bayesian Estimation of Ancestral States},
  author={Huelsenbeck, John P. and Bollback, Jonathan P.},
  journal={Systematic Biology},
  volume={50},
  number={3},
  pages={351--366},
  year={2001},
  doi={10.1080/10635150119871}
}

@article{yang2007paml,
  title={{PAML} 4: Phylogenetic Analysis by Maximum Likelihood},
  author={Yang, Ziheng},
  journal={Molecular Biology and Evolution},
  volume={24},
  number={8},
  pages={1586--1591},
  year={2007},
  doi={10.1093/molbev/msm088}
}

@article{guindon2010phyml,
  title={New Algorithms and Methods to Estimate Maximum-Likelihood Phylogenies: Assessing the Performance of {PhyML} 3.0},
  author={Guindon, St{\'e}phane and Dufayard, Jean-Fran{\c{c}}ois and Lefort, Vincent and Anisimova, Maria and Hordijk, Wim and Gascuel, Olivier},
  journal={Systematic Biology},
  volume={59},
  number={3},
  pages={307--321},
  year={2010},
  doi={10.1093/sysbio/syq010}
}

@article{nguyen2015iqtree,
  title={{IQ-TREE}: A Fast and Effective Stochastic Algorithm for Estimating Maximum-Likelihood Phylogenies},
  author={Nguyen, Lam-Tung and Schmidt, Heiko A. and von Haeseler, Arndt and Minh, Bui Quang},
  journal={Molecular Biology and Evolution},
  volume={32},
  number={1},
  pages={268--274},
  year={2015},
  doi={10.1093/molbev/msu300}
}

@article{katoh2013mafft,
  title={MAFFT Multiple Sequence Alignment Software Version 7: Improvements in Performance and Usability},
  author={Katoh, Kazutaka and Standley, Daron M.},
  journal={Molecular Biology and Evolution},
  volume={30},
  number={4},
  pages={772--780},
  year={2013},
  doi={10.1093/molbev/mst010}
}

@article{price2010fasttree,
  title={FastTree 2: Approximately Maximum-Likelihood Trees for Large Alignments},
  author={Price, Morgan N. and Dehal, Paramvir S. and Arkin, Adam P.},
  journal={PLOS ONE},
  volume={5},
  number={3},
  pages={e9490},
  year={2010},
  doi={10.1371/journal.pone.0009490}
}

@article{oliva2019ambiguity,
  title={Accounting for ambiguity in ancestral sequence reconstruction},
  author={Oliva, A. and Pulicani, S. and Lefort, V. and Brehelin, L. and Gascuel, O. and Guindon, S.},
  journal={Bioinformatics},
  volume={35},
  number={21},
  pages={4290--4297},
  year={2019},
  doi={10.1093/bioinformatics/btz249}
}

@article{Steinegger2017,
  author  = {Steinegger, Martin and S{\"o}ding, Johannes},
  title   = {MMseqs2 enables sensitive protein sequence searching for the analysis of massive data sets},
  journal = {Nature Biotechnology},
  year    = {2017},
  doi     = {10.1038/nbt.3988}
}

@article{10.1093/sysbio/22.3.240,
    author = {Felsenstein, Joseph},
    title = {Maximum Likelihood and Minimum-Steps Methods for Estimating Evolutionary Trees from Data on Discrete Characters},
    journal = {Systematic Biology},
    volume = {22},
    number = {3},
    pages = {240-249},
    year = {1973},
    month = {09},
    issn = {1063-5157},
    doi = {10.1093/sysbio/22.3.240},
    url = {https://doi.org/10.1093/sysbio/22.3.240},
    eprint = {https://academic.oup.com/sysbio/article-pdf/22/3/240/4741566/22-3-240.pdf},
}
}

\newpage


\appendix

\section{Technical appendices and supplementary material}

\subsection{ASR}
\label{app:classical_asr}

\paragraph{Classical ASR likelihood with among-site rate variation.}
Classical protein ASR models evolution at each aligned site as a continuous-time Markov chain over amino-acid states. For a tree \(T=(V,E)\) with leaves \(L\), internal nodes \(I=V\setminus L\), root \(r\), branch lengths \(t_{uv}\), and amino-acid rate matrix \(Q\), the site-wise likelihood is
\[
p(x_{L,j}\mid T,Q)
=
\sum_{h_{I,j}}
\pi(h_{r,j})
\prod_{(u,v)\in E}
\left[e^{t_{uv}Q}\right]_{h_{u,j},h_{v,j}} .
\]
Here \(x_{L,j}\) are the observed leaf states at alignment site \(j\), \(h_{I,j}\) are the unobserved internal states, \(\pi\) is the equilibrium distribution of \(Q\), and \(\left[e^{t_{uv}Q}\right]_{a,b}\) is the transition probability from amino acid \(a\) to amino acid \(b\) along branch \((u,v)\).

To account for among-site rate variation, the rate matrix is usually rescaled by a latent site-specific rate multiplier \(\lambda_j\). In practice, \(\lambda_j\) is commonly integrated out using a \(K\)-category discrete approximation to a Gamma distribution with shape parameter \(\alpha\). Let \(\lambda_k\) and \(w_k\) denote the rate and probability weight of category \(k\). The likelihood at site \(j\) becomes
\[
p(x_{L,j}\mid T,Q,\alpha)
=
\sum_{k=1}^{K}
w_k
\sum_{h_{I,j}}
\pi(h_{r,j})
\prod_{(u,v)\in E}
\left[e^{t_{uv}\lambda_k Q}\right]_{h_{u,j},h_{v,j}} .
\]
The dependence on \(\alpha\) is implicit through the discrete-Gamma category rates \(\lambda_k(\alpha)\) and weights \(w_k(\alpha)\). Assuming conditional independence across aligned sites, the full alignment likelihood factorizes as
\[
p(x_L\mid T,Q,\alpha)
=
\prod_{j=1}^{L_{\mathrm{MSA}}}
p(x_{L,j}\mid T,Q,\alpha).
\]

The Gamma rate categories are phenomenological: they improve likelihood by allowing some columns to evolve faster than others, but they do not explicitly model the structural or functional causes of that variation. The standard formulation also treats alignment columns independently and assumes that substitutions occur on a fixed coordinate system. Insertions and deletions are therefore usually handled before or outside the substitution model, through the construction of the MSA or through separate indel models.

\subsection{Model Architecture Details}
\label{app:model-architecture}

The main text describes the paired edit-flow transformer architecture. Here we specify the implementation details needed for reproducibility. L\ae{}rad operates on ungapped protein-token sequences, while alignments are retained as supervision and projection structures. Gap symbols are not emitted as amino-acid tokens; substitutions, insertions, and deletions are represented as separate edit operations.

Branch distances are provided to the model as an ordered two-dimensional condition,
\begin{equation}
\label{eq:branch_condition_app}
c_a =
\left(\frac{d_a}{\bar d}, \frac{d_b}{\bar d}\right),
\qquad
c_b =
\left(\frac{d_b}{\bar d}, \frac{d_a}{\bar d}\right),
\end{equation}
where \(\bar d\) is the dataset-level branch-budget scale. The first coordinate is the active child-to-parent budget, and the second is the paired-child budget.

The shared transformer trunk is followed by separate directional output heads. The heads predict substitution, insertion, and deletion rates; substitution and insertion heads also emit amino-acid distributions, while deletion is represented only by rate mass. During ASR inference, the child-to-parent head is used to decode observed children toward their inferred parent.

Training bridge states are sampled in aligned coordinates, but the transformer sees only gap-stripped protein sequences. Predicted rates and hidden states are projected back to aligned coordinates for route-comparison losses, same-LCA representation pooling, and site-wise evaluation.

\subsection{Batch Construction and Exact-LCA Grouping}
\label{app:batch-construction}

Training examples are directed leaf-pair ASR records. Each record contains two aligned descendants \((x_a,x_b)\), their exact pairwise LCA in the family tree, and branch edit distances \((d_a,d_b)\) from each descendant to that LCA. A nominal batch size \(B\) counts leaf-pair records, not individual sequence views; each batch therefore contains \(2B\) descendant sequences internally.

Mini-batches are sampled by exact-LCA groups. The grouping key is the family identifier together with the exact internal LCA node, so records from different trees cannot be grouped by accident. The sampler repeatedly selects an exact-LCA group and draws up to \(G\) records from it until the minibatch contains \(B\) records. In the reported experiments, \(G=3\), and at least two records from the same exact-LCA group are required for the same-LCA consistency loss.

The exact-LCA restriction means that records are grouped only when their pairwise LCA is the same internal node in the same family tree. Records that merely share a more distant ancestor, or whose LCAs are nested along the same lineage, are not used together for the group loss. Thus, the group term compares alternative descendant-pair views of the same ancestral node without imposing a global tree-wide collapse.

Sampling also uses a fixed mixture of the full eligible record pool and a hard pool enriched for low-identity pairs, high edit burden, or stronger indel signal. This mixture is constant throughout training and prevents the model from spending most updates on near-identical descendant pairs.

\subsection{Training hyperparameters}
\label{app:training_hyperparameters}

All reported L\ae{}rad models are trained with the same single-stage objective and optimizer schedule unless otherwise stated. We use batch size 48, a maximum of 80{,}000 optimization steps, AdamW-style weight decay \(10^{-2}\), initial learning rate \(8\times10^{-5}\), minimum learning rate \(10^{-5}\), gradient clipping at 1.0, and a 1{,}500-step warmup followed by learning-rate decay.

Bridge times are sampled with additional probability mass near the expected ancestral point \(\tau\). Exact-LCA grouping is enforced during batching: up to three records from the same exact-LCA group may be drawn into a minibatch, and at least two such records are required for the group-consistency loss.

The Base model required approximately 5--7 hours on a single NVIDIA A100 GPU. The Small model required approximately 1--2 hours on a single A100, while Tiny and Nano each required approximately 1--2 hours on a single NVIDIA RTX 4090.

An exploratory larger model with a 650M-parameter ESM-2 backbone required at least 30 GPU-hours on a single NVIDIA A100, was trained for 180{,}000 steps, and used a smaller minibatch size of 16 sequences. Both training and inference became substantially more expensive at this scale, illustrating the computational cost of scaling the model.

\subsection{Training Objective Details}
\label{app:training-objective}

The active objective is
\[
\mathcal{L}
=
w_{\mathrm{base}}\mathcal{L}_{\mathrm{Bregman}}
+
w_{\mathrm{ancestor}}\mathcal{L}_{\mathrm{ancestor}}
+
w_{\mathrm{group}}\mathcal{L}_{\mathrm{group}}.
\]
Earlier auxiliary losses for differentiable budget matching, site localization, operation-composition calibration, indel-excess control, rollout, mirror consistency, and staged curricula are disabled in the reported experiments. Branch distances are used for input conditioning, for locating \(\tau\), and for inference-time candidate selection.

\paragraph{Bregman edit-flow loss.}
Following Edit Flows~\cite{havasi2025editflowsflowmatching}, the base term trains a continuous-time edit-rate field by penalizing total predicted edit mass while rewarding rates assigned to edits that move a sampled bridge state toward the target endpoint.

For one directed route \(r:s\rightarrow t\), let \(z^0\) be the aligned source, \(z^1\) the aligned target, and
\[
z_t \sim p_t(\cdot\mid z^0,z^1)
\]
the sampled aligned bridge state. The model is evaluated on the gap-stripped version of \(z_t\), and its predicted rates are projected back to aligned coordinates. Let
\[
\hat u_{\theta,r}(i,o,a\mid z_t,t,c_r)
\]
denote the projected rate for operation \(o\in\{\mathrm{ins},\mathrm{sub},\mathrm{del}\}\) at aligned position \(i\), with emitted residue \(a\) for insertion and substitution. Deletion has no emitted residue.

Define \(\mathcal{A}(z_t,z^1)\) as the set of aligned one-step edits that move \(z_t\) toward \(z^1\):
\[
\mathcal{A}(z_t,z^1)
=
\{(i,\mathrm{ins},z^1_i): z_{t,i}=\mathrm{gap},\, z^1_i\neq \mathrm{gap}\}
\]
\[
\cup\,
\{(i,\mathrm{sub},z^1_i): z_{t,i}\neq \mathrm{gap},\, z^1_i\neq \mathrm{gap},\, z_{t,i}\neq z^1_i\}
\]
\[
\cup\,
\{(i,\mathrm{del}): z_{t,i}\neq \mathrm{gap},\, z^1_i=\mathrm{gap}\}.
\]
Let
\[
U_{\theta,r}(z_t,t,c_r)
=
\sum_j
\left(
\lambda^{\mathrm{ins}}_{\theta,r,j}
+
\lambda^{\mathrm{sub}}_{\theta,r,j}
+
\lambda^{\mathrm{del}}_{\theta,r,j}
\right)
\]
be the total predicted edit rate. With bridge scheduler \(\kappa(t)\), define
\[
\beta(t)=\frac{\dot{\kappa}(t)}{1-\kappa(t)}.
\]
In all reported experiments we use the linear scheduler \(\kappa(t)=t\).
The coefficient \(\dot{\kappa}(t)/(1-\kappa(t))\) is clipped for numerical stability.

The one-direction Bregman loss used in implementation is
\[
\ell_r(z_t,z^1,t,c_r)
=
\frac{1}{N(z_t,z^1)}
\left[
U_{\theta,r}(z_t,t,c_r)
-
\beta(t)
\sum_{a\in\mathcal{A}(z_t,z^1)}
\log \hat u_{\theta,r}(a\mid z_t,t,c_r)
\right],
\]
where \(N(z_t,z^1)=\max(|\mathcal{A}(z_t,z^1)|,1)\).

For a pair \((x_a,x_b)\), the forward component uses \(z^0=x_a\), \(z^1=x_b\), and time \(t\). The reverse component uses \(z^0=x_b\), \(z^1=x_a\), and time \(1-t\). Bridge times are sampled from a two-component mixture. Let \(p_\tau\in[0,1]\) denote the probability of drawing a near-ancestor bridge time. With probability \(1-p_\tau\), \(t\sim\mathrm{Uniform}(0,1)\); with probability \(p_\tau\), \(t\sim\mathrm{clip}(\mathcal{N}(\tau,\sigma_\tau^2),0,1)\). Thus \(p_\tau\) controls how often training examples are sampled near the expected ancestor, while \(\sigma_\tau\) controls the width of that near-\(\tau\) sampling window.

The same near-\(\tau\) scale is used to reweight the per-example Bregman losses:
\[
\omega_\tau(t)
=
1+
\alpha_\tau
\exp\left(
-\frac{1}{2}
\left(\frac{t-\tau}{\sigma_\tau}\right)^2
\right).
\]
Here \(\alpha_\tau\) controls the strength of the extra near-ancestor loss weight. Thus \(\omega_\tau(t)\to 1\) far from \(\tau\), while a sample exactly at \(\tau\) receives weight \(1+\alpha_\tau\). In the reported configuration, \(p_\tau=0.7\), \(\alpha_\tau=1.0\), and \(\sigma_\tau=0.08\).

This does not restrict training to the ancestral region: the uniform mixture component still covers the full bridge, and all examples retain nonzero weight. The near-\(\tau\) sampling and weighting only increase the density and contribution of bridge states near the point used for ASR inference. The directional losses are weighted across examples in each minibatch:
\[
\mathcal{L}_{a\rightarrow b}
=
\frac{\sum_n \omega_\tau(t_n)\ell_{a\rightarrow b}^{(n)}}{\sum_n \omega_\tau(t_n)},
\qquad
\mathcal{L}_{b\rightarrow a}
=
\frac{\sum_n \omega_\tau(t_n)\ell_{b\rightarrow a}^{(n)}}{\sum_n \omega_\tau(t_n)}.
\]
The bidirectional Bregman term is
\[
\mathcal{L}_{\mathrm{Bregman}}
=
\frac{1}{2}
\left(
\mathcal{L}_{a\rightarrow b}
+
\mathcal{L}_{b\rightarrow a}
\right).
\]

\paragraph{Ancestor latent-alignment loss.}
The ancestor term directly aligns the two route representations near \(\tau\). Hidden states from the two sampled routes are projected back to aligned coordinates. The loss is computed only for examples with \(|t-\tau|\leq r_{\mathrm{ancestor}}\) and over valid shared non-gap positions. For projected hidden states \(h^{a\rightarrow b}_{i}\) and \(h^{b\rightarrow a}_{i}\),
\[
\mathcal{L}_{\mathrm{ancestor}}
=
\lambda_{\cos}^{\mathrm{anc}}
\left(1-\cos(h^{a\rightarrow b}_{i},h^{b\rightarrow a}_{i})\right)
+
\lambda_{\ell_2}^{\mathrm{anc}}
\sqrt{
\frac{1}{d}
\left\|
h^{a\rightarrow b}_{i}
-
h^{b\rightarrow a}_{i}
\right\|_2^2
},
\]
averaged over valid near-\(\tau\) positions. This term does not use true ancestral sequences; it only asks the two descendant-conditioned routes to induce compatible latent states near the expected ancestor.

\paragraph{Exact-LCA group-consistency loss.}
The group term is applied only to records in the same minibatch whose exact pairwise LCA key is identical. For each record, the forward and reverse near-\(\tau\) hidden states are mean-pooled over valid shared non-gap aligned positions. These pooled route representations are treated as candidate representations of the same ancestral node. For two candidate representations \(r_i\) and \(r_j\) from different records with the same exact LCA,
\[
\mathcal{L}_{\mathrm{group}}(i,j)
=
\lambda_{\cos}^{\mathrm{group}}
\left(1-\cos(r_i,r_j)\right)
+
\lambda_{\ell_2}^{\mathrm{group}}
\sqrt{
\frac{1}{d}
\|r_i-r_j\|_2^2
}.
\]
The group loss is averaged over valid same-LCA candidate pairs, excluding comparisons between the two directions of the same record. It is not applied to arbitrary records from the same family or to nested but non-identical ancestral nodes.

In the reported configuration, \(w_\text{base}=w_\text{ancestor}=w_\text{group}=1.0\), 

\(\lambda^\text{anc}_\text{cos}=\lambda^\text{anc}_{\ell 2}=1.0\), 

\(\lambda^\text{group}_\text{cos}=1.0\), and 

\(\lambda^\text{group}_{\ell 2}=0.3\).

\subsection{Inference Procedure and Candidate Scoring}
\label{app:inference}
\label{app:inference_score}

Inference proceeds bottom-up on a binary tree. Leaves are initialized with their observed ungapped sequences. Once both children of an internal node have been populated, L\ae{}rad decodes candidate parent sequences from each child using the ordered branch conditions in Eq.~\ref{eq:branch_condition}.

For children \((x_a,x_b)\), decoding produces two directional candidate sets,
\[
\mathcal{C}_a=\{s_a^{(1)},\ldots,s_a^{(m)}\},
\qquad
\mathcal{C}_b=\{s_b^{(1)},\ldots,s_b^{(n)}\}.
\]
Candidates are retained when they occur near the target ancestral progress point or when their edit distance from the source child is close to the corresponding branch budget. L\ae{}rad first selects a compatible directional pair \((s_a,s_b)\), then constructs a consensus merge \(s_{\mathrm{merge}}\), and finally chooses among \(s_a\), \(s_b\), and \(s_{\mathrm{merge}}\).

Let \(\delta(\cdot,\cdot)\) denote Levenshtein edit distance. For a candidate ancestor \(s\), the branch-budget residual and parsimony terms are
\[
B(s)
=
|\delta(s,x_a)-d_a|
+
|\delta(s,x_b)-d_b|,
\qquad
P(s)
=
\delta(s,x_a)+\delta(s,x_b).
\]
For a directional pair \((s_a,s_b)\), the disagreement and model-support terms are
\[
D(s_a,s_b)=\delta(s_a,s_b),
\]
\[
M(s_a,s_b)
=
m_\theta(s_a\mid x_a,x_b,c_a)
+
m_\theta(s_b\mid x_b,x_a,c_b),
\]
where \(m_\theta\) is the stored average step score or log-probability of the decoded trajectory. Larger \(M\) indicates stronger model support.

The remaining fixed regularizers are
\[
E(s)
=
\max(0,|\delta(s,x_a)-d_a|-\epsilon_a)
+
\max(0,|\delta(s,x_b)-d_b|-\epsilon_b),
\]
\[
T(s)
=
\max(0,\delta(s,x_a)+\delta(s,x_b)-d_a-d_b-\epsilon_{ab}),
\]
and
\[
R(s)=|\rho(s)-\rho^\star|.
\]
Here \(E\) penalizes branch-budget residuals outside tolerance, \(T\) penalizes total-budget excess, and \(R\) penalizes off-target trajectory progress. The operation-profile penalty \(O(s)\) discourages unsupported insertion, deletion, and length-change profiles.

Inference uses two consecutive scoring steps. First, L\ae{}rad selects a compatible directional pair
\[
(s_a^*,s_b^*)
=
\arg\min_{(s_a,s_b)\in\mathcal{C}_a\times\mathcal{C}_b}
S_{\mathrm{pair}}(s_a,s_b).
\]
The pair score combines source-specific branch-budget residuals,
\[
|\delta(s_a,x_a)-d_a|+|\delta(s_b,x_b)-d_b|,
\]
cross-budget residuals,
\[
|\delta(s_a,x_b)-d_b|+|\delta(s_b,x_a)-d_a|,
\]
pair disagreement \(D(s_a,s_b)=\delta(s_a,s_b)\), average parsimony and operation-profile penalties, progress error, and the summed model-support term
\(M_{\mathrm{pair}}=m_\theta(s_a)+m_\theta(s_b)\), which enters with a negative sign.

After selecting \((s_a^*,s_b^*)\), L\ae{}rad constructs a consensus merge \(s_{\mathrm{merge}}\) and selects
\[
\hat{x}_v
=
\arg\min_{s\in\{s_a^*,s_b^*,s_{\mathrm{merge}}\}}
S_{\mathrm{final}}(s).
\]
The final score uses the single-candidate branch-budget residual \(B(s)\), parsimony \(P(s)\), budget-excess, total-budget-excess, operation-profile, and progress penalties. Pair-level disagreement and model support are inherited from the selected directional pair. If any of the three final candidates satisfy both branch-budget tolerances, candidates outside tolerance are discarded before applying \(S_{\mathrm{final}}\). The numerical weights used in \(S_{\mathrm{pair}}\) and \(S_{\mathrm{final}}\) are fixed across all reported runs and are provided in the released evaluation configuration. We report the scoring components here because the implementation applies them in two stages: pair selection over \(\mathcal{C}_a\times\mathcal{C}_b\), followed by final selection among \(s_a^*\), \(s_b^*\), and \(s_{\mathrm{merge}}\).

The branch-budget tolerance used for candidate filtering is
\[
\epsilon(d)=\max(1,0.1d).
\]
Candidate states are retained if their decoded trajectory progress is close to the target progress or if their edit distance from the source child lies within the branch-budget tolerance. For the \(x_a\)-derived route, the target progress is \(\rho^\star_a=\tau\); for the \(x_b\)-derived route, it is \(\rho^\star_b=1-\tau\). For the merged candidate, progress and target progress are averaged from the selected directional pair.

The reported inference uses \(n_{\mathrm{steps}}=25\), \(N=1\) sampled trajectory per child, and a candidate progress window of \(0.15\) for the released J-protein evaluation. The sampler is a discrete tau-leaping approximation. At each progress step with adaptive step size \(h_t\), insertion events are sampled with probability \(1-\exp(-h_t\lambda^{\mathrm{ins}})\), while substitution/deletion events are sampled with probability \(1-\exp[-h_t(\lambda^{\mathrm{sub}}+\lambda^{\mathrm{del}})]\). Conditional on a substitution/deletion event, deletion is chosen with probability \(\lambda^{\mathrm{del}}/(\lambda^{\mathrm{sub}}+\lambda^{\mathrm{del}})\); otherwise a substitution residue is sampled from \(q^{\mathrm{sub}}\). Inserted residues are sampled from \(q^{\mathrm{ins}}\).

\subsection{Edit operations and edit-flow parameterization}
\label{app:edit_operations}

Following the edit-flow formulation of Havasi et al.~\cite{havasi2025editflowsflowmatching}, L\ae{}rad models sequence evolution through elementary insertion, deletion, and substitution operations. Let \(x=(x_1,\dots,x_n)\in\mathcal{V}^n\) be a sequence over vocabulary \(\mathcal{V}\). For a position \(i\) and symbol \(v\in\mathcal{V}\),
\[
    \operatorname{ins}(x,i,v)
    =
    (x_1,\ldots,x_i,v,x_{i+1},\ldots,x_n),
\]
\[
    \operatorname{del}(x,i)
    =
    (x_1,\ldots,x_{i-1},x_{i+1},\ldots,x_n),
\]
\[
    \operatorname{sub}(x,i,v)
    =
    (x_1,\ldots,x_{i-1},v,x_{i+1},\ldots,x_n).
\]
Insertion adds a token after position \(i\), deletion removes token \(i\), and substitution replaces token \(i\) by \(v\).

The continuous-time edit field separates operation intensity from emitted-token identity. Here \(c\) denotes the ordered branch condition from Eq.~\ref{eq:branch_condition}, so the edit field is conditioned on both the active child-to-parent budget and the paired-child budget:
\[
    u_t^\theta(\operatorname{ins}(x,i,v)\mid x,c)
    =
    \lambda^{\mathrm{ins}}_{\theta,t,i}(x,c)\,
    q^{\mathrm{ins}}_{\theta,t,i}(v\mid x,c),
\]
\[
    u_t^\theta(\operatorname{del}(x,i)\mid x,c)
    =
    \lambda^{\mathrm{del}}_{\theta,t,i}(x,c),
\]
\[
    u_t^\theta(\operatorname{sub}(x,i,v)\mid x,c)
    =
    \lambda^{\mathrm{sub}}_{\theta,t,i}(x,c)\,
    q^{\mathrm{sub}}_{\theta,t,i}(v\mid x,c).
\]
Thus, insertions and substitutions factor into a position-specific rate \(\lambda\) and a categorical distribution \(q\) over emitted amino acids, while deletions require only a position-specific rate. In implementation, the three rate channels are ordered as insertion, substitution, and deletion, and the insertion/substitution distributions are masked to the standard amino-acid alphabet rather than emitting gap tokens.

The bridge endpoints do not restrict the emitted residue vocabulary. Although the sampled aligned bridge state is constructed between two descendants, the substitution and insertion heads output distributions over the full amino-acid alphabet, not only over residues present in the two focal descendants. Therefore L\ae{}rad can in principle propose an ancestral residue absent from both children. Such out-of-pair recovery is not identifiable from a single descendant pair alone; it must come from the learned sequence-context prior, paired cross-attention, and agreement across other descendant pairs that share the same ancestral node. Candidate scoring can then retain such residues if they improve the decoded ancestor under the model and branch-consistency terms.
\subsection{Dataset-specific filtering details}
\label{app:data_filtering}

All datasets are processed into family-level ASR artifacts, with source-specific filters.

\paragraph{FPbase fluorescent proteins.}
For FPbase, sequences are retained only if they are non-empty, begin with methionine, have length 50--2000, and contain only standard amino-acid residues. We remove low-complexity sequences with single-residue fraction above 0.40 or homopolymer runs longer than 12, and discard duplicated protein names. Retained sequences are clustered into families with MMseqs2~\cite{Steinegger2017} using minimum sequence identity 0.50 and coverage 0.80.

\paragraph{eggNOG orthologous groups.}
For eggNOG, we retain orthologous groups that contain both a raw multiple sequence alignment and a matching precomputed eggNOG tree. In the reported preprocessing, retained groups have 2--128 aligned members, equal aligned sequence lengths, exact agreement between alignment and tree leaves, and alignment length at most 384 columns.

\paragraph{Bacteriophage J proteins.}
For the phage J-protein dataset, we use the provided full-length DRYAD alignments and matching FastTree~\cite{price2010fasttree} topologies. Non-standard residues are converted to gaps so that all sequences can be encoded by the ASR alphabet. We require equal aligned sequence lengths and exact agreement between tree leaves and alignment records. In the current artifacts, the retained full-length ID95 family contains 70 sequences and the retained full-length ID99 family contains 1207 sequences.

\subsection{Phylogenetic tree construction and pair-generation details}
\label{app:tree_pair_generation}

L\ae{}rad is trained from leaf-pair relationships rather than experimentally known ancestral sequences. For each family, we require a topology, a multiple sequence alignment, and branch distances expressed in edit-operation units.

\paragraph{Tree sources and construction.}
For FPbase, phylogenetic trees are inferred because none are provided by the source database. We first align each family with MAFFT~\cite{katoh2013mafft}. Families with at least four leaves are assigned maximum-likelihood topologies with RAxML-NG~\cite{10.1093/bioinformatics/btz305} under an LG+G8+F model; smaller families use trivial or neighbor-joining fallback trees. For eggNOG, we use the provided eggNOG topology with the raw family alignment. For the phage J-protein dataset, we use the provided DRYAD/FastTree topology with the full-length J-protein alignment.

\paragraph{Edit-budget construction.}
To convert tree edges into edit-operation budgets, we estimate internal node states with Fitch parsimony and count aligned substitutions, deletions, and insertions along each parent-child edge. These Fitch states are used only to assign branch edit distances; they are not used as supervised ancestral targets for L\ae{}rad.

\paragraph{ASR pair records.}
Training records are constructed from directed leaf pairs. For each pair, we compute the exact lowest common ancestor, the edit distance from each leaf to that ancestor, the pairwise aligned sequences, percent identity, and total edit count. These records define the branch-conditioned ASR problem used during training.

\subsection{Metric definitions}
\label{app:metric_definitions}

All Pearson correlations are computed over finite aligned-site entries only. If fewer than three valid sites are available, or either vector has zero variance, the metric is undefined.

\paragraph{Observed edit correlation.}
The J-protein benchmark has no known internal ancestral sequences, so exact ancestral accuracy cannot be computed. We therefore evaluate whether inferred evolutionary changes fall at sites that are empirically variable among extant sequences. All site-wise metrics are computed in the fixed coordinate system of the J-protein MSA.

First, an empirical leaf-level operation profile is constructed by comparing each aligned leaf sequence to a fixed aligned reference leaf. In all reported J-protein experiments, the fixed reference leaf is \texttt{TIPJ\_LAMBD/1-1132}. The aligned sequence of this leaf defines the MSA coordinate system used both to construct the empirical leaf-vs-reference operation profile and to project ungapped L\ae{}rad reconstructions back to aligned sites before scoring.

For an ungapped L\ae{}rad prediction, we globally align the prediction to the ungapped reference sequence using Needleman--Wunsch dynamic programming with match score \(+1\), mismatch score \(-1\), and gap score \(-1\). We then traverse the original aligned reference MSA row: non-gap reference columns receive the prediction residue aligned to that reference residue, or a gap if no prediction residue is aligned; reference-gap columns are assigned gaps in the projected prediction. This preserves the original MSA length and does not introduce new columns for insertions relative to the reference. Operation counts are then computed in the shared MSA coordinate system: a residue-to-different-residue transition contributes one substitution count, a residue-to-gap transition contributes one deletion count, and a gap-to-residue transition contributes one insertion count.

Second, each method is converted into an inferred branch-operation profile. For every scored tree edge, parent and child states are compared in the same MSA coordinate system. L\ae{}rad outputs ungapped sequences, so its reconstructions are projected back to the reference-aligned MSA coordinates before scoring. Counts are assigned as
\[
\text{residue}\to\text{different residue} =\mathrm{substitution},
\]
\[
\text{residue}\to\text{gap}=\mathrm{deletion},
\]
\[
\text{gap}\to\text{residue}=\mathrm{insertion}
\]
Accumulating over scored edges gives \(M^{(m)}_{\mathrm{sub},i}\), \(M^{(m)}_{\mathrm{del},i}\), and \(M^{(m)}_{\mathrm{ins},i}\) for method \(m\).

The observed edit correlation is then
\[
\rho_{\mathrm{edit}}^{(m)}
=
\operatorname{Pearson}
\left(
\left(M^{(m)}_{\mathrm{sub},i}+M^{(m)}_{\mathrm{del},i}+M^{(m)}_{\mathrm{ins},i}\right)_{i=1}^{L},
\left(O_{\mathrm{sub},i}+O_{\mathrm{del},i}+O_{\mathrm{ins},i}\right)_{i=1}^{L}
\right).
\]
This is a proxy for localization of evolutionary change, not direct ancestral accuracy.

\paragraph{Observed indel correlation.}
Observed indel correlation uses the same site-wise construction but restricts both profiles to deletion and insertion counts:
\[
\rho_{\mathrm{indel}}^{(m)}
=
\operatorname{Pearson}
\left(
\left(M^{(m)}_{\mathrm{del},i}+M^{(m)}_{\mathrm{ins},i}\right)_{i=1}^{L},
\left(O_{\mathrm{del},i}+O_{\mathrm{ins},i}\right)_{i=1}^{L}
\right).
\]
For methods that reconstruct aligned states rather than explicit indel events, gap/residue transitions are scored post hoc as alignment-state changes. These are comparable site-wise signals, but they should not be interpreted as native indel-event reconstructions.

\paragraph{Normalized budget error.}
For each scored tree edge \(e\), let \(K_e^{(m)}\) be the total number of inferred edit operations on that edge, and let \(d_e\) be the tree-implied branch edit budget. The normalized budget error is
\[
R_{\mathrm{norm}}^{(m)}
=
\frac{1}{|E_s|}
\sum_{e\in E_s}
\frac{|K_e^{(m)}-d_e|}{\max(d_e,1)}.
\]
Lower values indicate better agreement between inferred edit counts and branch budgets.

\paragraph{Fluorescent-protein exact accuracy.}
For the fluorescent-protein benchmark, known internal ancestral sequences are available. Predicted internal sequences are projected to the known ancestral sequence coordinates. A site is correct when the projected predicted residue equals the known ancestral residue. The exact accuracy is
\[
A_{\mathrm{exact}}^{(m)}
=
100\,
\frac{\sum_h C_h^{(m)}}{\sum_h L_h},
\]
where \(C_h^{(m)}\) is the number of correctly inferred sites for internal node \(h\), and \(L_h\) is the length of the known ancestral sequence. The number of incorrectly inferred sites is
\[
N_{\mathrm{err}}^{(m)}
=
\sum_h \left(L_h-C_h^{(m)}\right).
\]

\subsection{Licenses}
\label{app:licenses}

This work relies on publicly available datasets and external software (including model architectures). We cite the original sources throughout the paper and summarize the corresponding licenses or access terms here. Our supplementary release redistributes only material that we are permitted to share directly; third-party raw assets must be obtained from their original sources.

\paragraph{Datasets.}
\begin{itemize}
    \item \textbf{FPbase fluorescent proteins}~\cite{lambert2016fpbase}. The \emph{FPbase website and source code} are released under a \href{https://creativecommons.org/licenses/by-sa/4.0/}{CC BY-SA 4.0} license, while the \emph{data contained in FPbase} are described by FPbase as free of copyright restrictions and available for non-commercial and commercial use with attribution to the original authors of the corresponding data.
    
    \item \textbf{eggNOG 5.0}~\cite{10.1093/nar/gky1085}. We use publicly accessible orthologous groups, alignments, and trees from eggNOG under its publicly provided access terms. We did not identify an explicit reuse license for the exact assets used here, and therefore do not redistribute raw eggNOG assets.
    
    \item \textbf{Bacteriophage J dataset / Dryad archive}~\cite{10.1111/evo.13586}. Dryad datasets are made publicly available under the \href{https://creativecommons.org/publicdomain/zero/1.0/}{CC0 1.0 Universal} dedication.
    
    \item \textbf{Experimental fluorescent-protein benchmark}~\cite{randall2016asrbenchmark}. The associated article is published under \href{https://creativecommons.org/licenses/by/4.0/}{CC BY 4.0}. The benchmark data are described by the authors as available upon request, so we cite the original source and do not redistribute the raw benchmark files.
\end{itemize}

\paragraph{External software.}
\begin{itemize}
    \item \textbf{PAML}~\cite{yang2007paml}: \href{https://www.gnu.org/licenses/gpl-3.0.en.html}{GNU GPL v3}.
    \item \textbf{PhyML}~\cite{guindon2010phyml,oliva2019ambiguity}: \href{https://www.gnu.org/licenses/gpl-3.0.en.html}{GNU GPL v3}.
    \item \textbf{IQ-TREE}~\cite{nguyen2015iqtree}: \href{https://www.gnu.org/licenses/old-licenses/gpl-2.0.en.html}{GNU GPL v2}.
     \item \textbf{ARPIP}~\cite{10.1093/sysbio/syac050}: \href{https://www.gnu.org/licenses/lgpl-3.0.html}{GNU Lesser General Public License v3.0 or later (LGPL-3.0-or-later)}.
    \item \textbf{AutoRegressiveASR}~\cite{10.1093/molbev/msaf070}. We did not identify an explicit software license on the public repository page consulted for this submission, so we cite the original paper and do not redistribute its code.
    \item \textbf{RaxML-NG}~\cite{10.1093/bioinformatics/btz305}: \href{https://www.gnu.org/licenses/agpl-3.0.html}{GNU Affero General Public License v3} 
    \item \textbf{MMseqs-2}~\cite{Steinegger2017}: \href{https://opensource.org/license/mit}{MIT License}
    \item \textbf{ESM-2}~\cite{esm2}. The public ESM-2 model card used in our workflow lists the \href{https://opensource.org/license/mit}{MIT License}.
\end{itemize}

\end{document}